\documentclass[aps,preprint,prb]{revtex4}
\usepackage{amssymb}
\usepackage{amsmath}
\usepackage{graphicx}
\usepackage{latexsym}
\usepackage{bm}

\begin{document}

\author{Sharif D. Kunikeev and David L. Freeman}
\affiliation{Department of Chemistry \\
University of Rhode Island \\
Kingston, RI 02881 }
\author{ J.D. Doll}
\affiliation{Department of Chemistry \\
Brown University \\
Providence, RI 02912}
\title{ {\Large Convergence Characteristics of the Cumulant Expansion for Fourier Path Integrals}}

\begin{abstract}
The cumulant representation of the Fourier path integral method is examined to determine the asymptotic convergence characteristics of the imaginary-time density matrix with respect to the number of path variables $N$ included.  It is proved that when the cumulant expansion is truncated at order $p$, the asymptotic convergence rate of the density matrix behaves like $N^{-(2p+1)}$.  The complex algebra associated with the proof is simplified by introducing a diagrammatic representation of the contributing terms along with an associated linked-cluster theorem.   The cumulant terms at each order are expanded in a series such that the the asymptotic convergence rate is maintained without the need to calculate the full cumulant at order $p$.  Using this truncated expansion of each cumulant at order $p$, the numerical cost in developing Fourier path integral expressions having convergence order $N^{-(2p+1)}$ is shown to be approximately linear in the number of required potential energy evaluations making the method promising for actual numerical implementation.
\end{abstract}

\maketitle

\section{Introduction}

The path integral method\cite{feynman_quantum_1965,berne_simulation_1986,doll_equilibrium_1990-1,ceperley_path_1995-1,kleinert_path_1995,chaichian_path_2001} has proved to be an important and useful computational vehicle for obtaining thermodynamic properties of interacting many-particle systems in the quantum domain. In all path integral approaches the usual classical degrees of freedom in a system are augmented by an infinite set of path variables that effectively describe the quantum fluctuations about the classical trajectories.  In actual simulations the infinite set of path variables is truncated, and an important issue is the rate of convergence to the exact quantum result as a function of the number of path variables actually included. 

In one approach\cite{schweizer_convenient_1981,berne_simulation_1986,ceperley_path_1995-1} to path integration the imaginary-time propagator is discretized in coordinate representation using a large set of intermediate coordinate states along with the Trotter approximation.\cite{trotter__1959}  The Trotter decomposition becomes increasingly accurate as the number of discretized path variables is increased.  The asymptotic convergence rate of this discretized version of path integrals is known\cite{ceperley_path_1995-1} to be $1/N^2$ where $N$ is the number of discretized points included.  For problems where pair potentials are adequate, the convergence of the discretized method can be enhanced by, for example, using more accurate pair propagators.\cite{ceperley_path_1995-1}

In this work we focus on an alternate path integral method\cite{freeman_mboxmonte_1984,freeman_quantum_1985,doll_equilibrium_1990-1} where the quantum paths are expanded in a Fourier series, and the integration over all paths is replaced by a Riemann integral with respect to the Fourier coefficients.  While exact results are obtained if the complete Fourier series containing an infinite set of terms is included, in practical applications, the number of coefficients included is truncated at $N$ terms.  In its primitive form this Fourier path integral method is known\cite{eleftheriou_asymptotic_1999-1} to converge asymptotically to exact results as $1/N$.  

To enhance the asymptotic convergence rate of the Fourier method, a set of useful approaches has been developed to include approximately the contributions from the coefficients excluded when the full Fourier series is truncated.  The first of these methods has been named ``partial averaging,''\cite{doll_mboxfourier_1985,coalson_partial_1986,doll_equilibrium_1990-1} because integrals over the high-order Fourier path variables, the ``tail integrals,'' are included in an average sense.  Because partial averaging requires the evaluation of the Gaussian transform of the interaction potentials associated with a particular problem, and because many interaction potentials used commonly in simulations do not have finite or readily available Gaussian transforms, alternative methods have also been introduced that circumvent the need for a Gaussian transform.  Among such methods are the gradient partial average method\cite{coalson_partial_1986,doll_equilibrium_1990-1} and the reweighted path integral technique.\cite{predescu_optimal_2002,predescu_numerical_2003-1,predescu_random_2003,predescu_energy_2003,predescu_heat_2003-1,predescu_asymptotic_2003}  The asymptotic convergence rates of the full partial average method\cite{predescu_asymptotic_2003} and the reweighted method\cite{predescu_optimal_2002} are both $1/N^3$, whereas the gradient partial average method converges as $1/N^2$.\cite{eleftheriou_asymptotic_1999-1}

In a previous publication\cite{kunikeev_numerical_2009} we considered a suggestion by Singer\cite{singer_use_1958} to fit the Lennard-Jones potential, which does not have a finite Gaussian transform, to a sum of two Gaussians, which have Gaussian transforms that are both analytic and finite.  In that previous work we showed the fit potential to be an accurate representation of a one-dimensional Lennard-Jones system, and we explored the asymptotic convergence characteristics using a variety of Fourier path integral methods.  In that work we found numerically, for the case studied, that the full partial average method reached its asymptotic limit more rapidly than the reweighted method.  Recalling that the reweighted method has the same asymptotic convergence rate as the full partial average method, the results indicated that the full partial average method can be advantageous.  

The partial average method can be understood\cite{coalson_partial_1986,doll_equilibrium_1990-1} as the first term in a cumulant expansion\cite{kubo_generalized_1962,coalson_cumulant_1989} with respect to the tail series of the quantum density matrix.  Motivated by the successful implementation of the partial average method using Gaussian fit potentials, in this work we explore the asymptotic convergence characteristics of the approach when cumulants are included to arbitrary rather than just first order.  While the cumulant expansion for Fourier path integrals has been examined through second order in previous work,\cite{coalson_cumulant_1989} the properties of the full expansion have not been discussed before.  The cumulant expansion by itself is not guaranteed to be convergent.  In the current work we ignore potential convergence issues with respect to cumulant order so that we can explore the asymptotic convergence characteristics of the cumulant expansion with respect to the number of included path variables $N$ when the cumulant expansion itself is truncated at an arbitrary but finite order $p$.
 The analysis is complex, because at each cumulant order beyond $p=1$, there are many terms having differing convergence characteristics with respect to $N$.  We find many of the resulting terms cancel in a way that allows us to show that the resulting asymptotic convergence rate is $N^{-(2p+1)}$.  The analysis required is simplified by using a combination of algebraic and diagrammatic methods.  The diagrammatic approach allows us to prove a linked-cluster theorem that tells us which of the large number of terms at each order in $p$ survive cancellation.  From the proved asymptotic convergence rate, the first-order cumulant (partial averaging) converges asymptotically as $1/N^3$, including the second-order cumulant enhances the asymptotic convergence rate to $1/N^5$ and so on.  The derivation of the $N^{-(2p+1)}$ asymptotic convergence rate for cumulants of order $p$ is a principal finding of the current work.

The derivations of the asymptotic convergence rates rely on certain expansions at each cumulant order.  We show that these expansions at each cumulant order, when truncated, have the same asymptotic convergence characteristics as the full cumulants.  Furthermore, the truncated expansions at each cumulant order enable the development of path integral approaches that scale approximately linearly in the number of required potential energy evaluations while retaining the $N^{-(2p+1)}$ asymptotic convergence rate.  This important finding implies the developments described in this work are potentially important from a numerical standpoint.

The contents of the remainder of this paper are as follows.  In the next section we present our theoretical developments.  Included is a review of the Fourier path integral method, the full partial average method and the cumulant expansion.  As examples we derive explicitly the asymptotic convergence rates of the first through third order cumulant terms.  We then present a diagrammatic representation of the cumulant expansion that enables the derivation of the asymptotic convergence rate at any finite order of cumulant truncation.  Much of the detailed examination of the convergence properties of the terms is given in two appendices.  In Section III we review our key findings and discuss the expected numerical work required to implement the higher-order cumulants in actual calculations.

\section{Theory}

\subsection{Fourier path integral representations of the density matrix and partial averaging} \label{sec:oned}

In this subsection we introduce the path integral representation of the quantum imaginary time density matrix, and demonstrate the utility of the Fourier representation of the paths in terms of a tail series.  Some of the discussion in the subsection can be found in previous literature,\cite{doll_equilibrium_1990-1,predescu_random_2003} but we find it useful to repeat the details for clarity and to establish the notation needed for the subsequent development.  For simplicity we assume a one-dimensional system with the extension to multidimensional systems to be presented with numerical examples in a separate publication.  

We consider the Fourier
representation of the Feynman-Kac formula for the quantum density matrix \cite{chaichian_path_2001}
\begin{eqnarray}
\rho (x,x^{\prime };\beta ) &=&\rho _{fp}(x,x^{\prime };\beta )K(x,x^{\prime
};\beta )  \notag \\
&=&\rho _{fp}(x,x^{\prime };\beta )\prod_{k=1}^{\infty }\left(
\int\limits_{-\infty }^{\infty }\frac{da_{k}}{\sqrt{2\pi }}\exp \left( -%
\frac{1}{2}a_{k}^{2}\right) \right) \exp \left( -\beta \overline{V}%
(x,x^{\prime },\{a_{k}\};\beta )\right)  \label{1.1}
\end{eqnarray}%
where $\rho _{fp}$ is the free-particle density matrix and $%
\beta =1/(k_{B}T)$ the inverse temperature with $k_B$ the Boltzmann constant.  We have used the notation
\begin{equation}
\overline{f} = \int_0^1 f[x(u)] \ du
\end{equation}
to represent an average of a function $f$ with respect to the ``imaginary time'' variable u.  In a similar manner, for a function $g$ of two or more variables, we can write
\begin{equation}
\overline{g} = \int_0^1 du_1 \int_0^1 du_2 \ldots \ g[x(u_1),y(u_2),\ldots].
\end{equation}
Using this notation the time average of the potential in Eq.(\ref{1.1}) is
\begin{equation}
\overline{V}(x,x^{\prime },\{a_{k}\};\beta ) =\int\limits_{0}^{1}du\,V%
\left[ x_{r}(u)+\sigma \vec{a}\cdot \vec{\Lambda}(u)\right] ,  \label{1.2}
\end{equation}
with
\begin{equation}
x_{r}(u)=x+(x^{\prime }-x)u,
\end{equation}
\begin{equation}
 \sigma =\sqrt{\frac{\hbar ^{2}\beta }{m}},
\end{equation}
\begin{equation}
\vec{a} =(a_{1},a_{2},\ldots ),
\end{equation}
\begin{equation}
\vec{\Lambda}(u)=(\Lambda _{1}(u),\Lambda _{2}(u),\ldots ),
\end{equation}
\begin{equation}
 \Lambda
_{k}(u)=\sqrt{2}\frac{\sin (\pi ku)}{\pi k},
\end{equation}
and $m$ is the particle's mass. The paths in Eq. (\ref{1.2}), starting at coordinate $x$ and
ending at coordinate $x^{\prime }$, are expanded in Fourier series, and each
path is parametrized by an infinite set of Fourier coefficients $%
(a_{1},a_{2},\ldots )$. In order to calculate integral over the paths in Eq. (%
\ref{1.1}), one needs to carry out an infinite dimensional integration over
Fourier coefficients $a_{1},a_{2},\ldots $. Because an infinite dimensional integration is not numerically feasible, we truncate to a finite dimensional integral.  The resulting finite dimensional integration
converges to the exact result as the number of integration variables
increases. To this end, we divide infinite dimensional integral Eq. (\ref%
{1.1}) into a finite dimensional integral over the first $N$ variables $\vec{%
a}_{N}=(a_{1},a_{2},\ldots ,a_{N})$ and the tail integration  (TI) over the
infinite dimensional tail $\vec{a}_{N+1}=(a_{N+1},a_{N+2},\ldots )$%
\begin{equation}
K(x,x^{\prime };\beta )=\prod_{k=1}^{N}\left( \int\limits_{-\infty }^{\infty
}\frac{da_{k}}{\sqrt{2\pi }}\exp \left( -\frac{1}{2}a_{k}^{2}\right) \right)
\left\langle \exp (-\beta \overline{V})\right\rangle _{TI}  \label{1.3}
\end{equation}%
where the TI denotes the tail integration
\[
\left\langle \exp (-\beta \overline{V})\right\rangle _{TI}
=\prod_{k=N+1}^{\infty }\left( \int\limits_{-\infty }^{\infty }\frac{da_{k}%
}{\sqrt{2\pi }}\exp \left( -\frac{1}{2}a_{k}^{2}\right) \right)
\]
\begin{equation}
 \times \exp \left( -\beta \int\limits_{0}^{1}du\,V\left[ x_{N}(u)+\sigma \vec{a}%
_{N+1}\cdot \vec{\Lambda}_{N+1}(u)\right] \right) . \label{eq:ti}
\end{equation}%
In Eq.(\ref{eq:ti}) we have introduced the notation
\begin{equation}
x_{N}(u)=x_{r}(u)+\vec{a}_{N}\cdot \vec{\Lambda}_{N}(u)=x_{r}(u)+%
\sum_{k=1}^{N}a_{k}\Lambda _{k}(u).
\end{equation}%
The TI variables appear in the argument of potential function linearly, which can be expressed as a scalar product 
\begin{equation}
\vec{a}_{N+1}\cdot \vec{\Lambda}_{N+1}(u)=\sum_{k=N+1}^{\infty }a_{k}\Lambda
_{k}(u). \label{eq:sp}
\end{equation}%

Equation (\ref{eq:sp}) suggests a possible change of integration variables to a set of projection
variables onto the vectors $\vec{\Lambda}_{N+1}(u)$, so that integration in
the subspace orthogonal to this projection subspace might be performed
trivially. The problem with a direct realization of this idea is that Eq. (%
\ref{1.1}) contains an integration over $u$, and we have a continuum set of
vectors $\vec{\Lambda}_{N+1}(u)$. However, we can use the linearity of the
integration operation allowing the order of integration in Eq.(\ref{eq:ti}) to be
changed. To understand how the change of integration order helps evaluate Eq. (\ref{eq:ti}), we expand the exponential function using a Taylor series%
\begin{equation}
\left\langle \exp (-\beta \overline{V})\right\rangle
_{TI}=1+\sum_{p=1}^{\infty }\frac{(-\beta )^{p}}{p!}\mu _{p} \label{1.4}%\equiv \exp (%
%\mathcal{V}_{cum}) 
\end{equation}%
where the TI of the $p^{th}$ power of $\overline{V}$ defines the $p^{th}$ order
moment%
\begin{eqnarray}
\mu _{p} &\equiv &\left\langle \left( \int\limits_{0}^{1}du\,V\left[
x_{N}(u)+\sigma \vec{a}_{N+1}\cdot \vec{\Lambda}_{N+1}(u)\right] \right)
^{p}\right\rangle _{TI}  \notag \\
&=&\int\limits_{0}^{1}\cdots \int\limits_{0}^{1}du_{1}\cdots
du_{p}\left\langle \prod\limits_{k=1}^{p}V\left[ x_{N}(u_{k})+\sigma \vec{a}%
_{N+1}\cdot \vec{\Lambda}_{N+1}(u_{k})\right] \right\rangle _{TI}.
\label{1.5}
\end{eqnarray}
In the second line of Eq. (\ref{1.5}) we have interchanged the order of
integration over $u$-variables and the TI variables $\vec{a}_{N+1}$. At a
fixed set of external time variables $\{u_{1},\ldots ,u_{p}\}$, we now have
a fixed set of vectors $\{\vec{\Lambda}_{N+1}(u_{1}),\ldots ,\vec{\Lambda}%
_{N+1}(u_{p})\}$ on which the vector $\vec{a}_{N+1}$ is projected. To proceed further, we recognize that in
general, the $\vec{\Lambda}$ vectors are not orthonormalized. It is useful
if we normalize the vectors and introduce non-orthogonal unit vectors 
\begin{equation}
\vec{g}_{k}\equiv \vec{\Lambda}_{N+1}(u_{k})/\sqrt{\varepsilon (u_{k})}%
,\quad k=1,2,\ldots ,p \label{eq:gk}
\end{equation}%
where square of the normalization factor%
\begin{equation}
\varepsilon (u_{k})=\gamma (u_{k},u_{k})\equiv \vec{\Lambda}%
_{N+1}(u_{k})\cdot \vec{\Lambda}_{N+1}(u_{k})=\sum_{n=N+1}^{\infty }\Lambda
_{n}^{2}(u_{k})
\end{equation}%
is a natural small parameter useful for much of the analysis found in the remainder of this work.  We write
\begin{equation}
%\lim_{N \rightarrow \infty}
\varepsilon (u_k) \sim \frac{1}{N},
\end{equation}
which, as explained and used in reference \onlinecite{kunikeev_numerical_2009}, is a
 shorthand notation for asymptotic behavior of
an integral with respect to $u$  of the
product of $\varepsilon (u)$ and a smooth function $f(u)$ ; i.e.
\begin{equation}
\int\limits_{0}^{1}du\,\varepsilon (u)f(u)=O\left( \frac{1}{N}\right).
\end{equation}%
%as $N\rightarrow \infty $.

The set of vectors $\{g_k\}$ defined in Eq. (\ref{eq:gk}) is normalized but not orthogonal.  It is useful to work with an orthogonal set that can by obtained from $\{g_k\}$ using Gramm-Schmidt orthogonalization procedure
\begin{equation}
\vec{g}_{k}=\sum_{i=1}^{k}\alpha _{ki}\vec{e}_{i},\qquad
\;\sum_{i=1}^{k}\alpha _{ki}^{2}=1,  \label{GS1}
\end{equation}%
where the vectors $\{\vec{e}_{1},\cdots ,%
\vec{e}_{p}\}$ are expanded to ensure orthogonality; i.e. $\vec{e}_{i}\cdot \vec{e}_{k}=\delta _{ik}$. The
coefficients $\alpha _{ki}$ and the vectors $\vec{e}_{i}$ can be easily
calculated using the standard recurrence relations, so that we obtain 
\begin{equation}
\vec{e}_{k}=\left( \vec{g}_{k}-\sum_{i=1}^{k-1}\alpha _{ki}\vec{e}%
_{i}\right) /\alpha _{kk},  \label{GS2}
\end{equation}%
where 
\begin{eqnarray}
\alpha _{ki} &=&\vec{g}_{k}\cdot \vec{e}_{i},\qquad i=1,\cdots ,k-1,
\label{GS3} \\
\alpha _{kk} &=&\sqrt{1-\sum_{i=1}^{k-1}(\vec{g}_{k}\cdot \vec{e}_{i})^{2}}
\label{GS4}
\end{eqnarray}%
Using these relations, we obtain, e.g., for the first three vectors 
\begin{eqnarray}
\vec{e}_{1} &=&\vec{g}_{1},\qquad \vec{e}_{2}=(\vec{g}_{2}-g_{21}\vec{g}%
_{1})/\sqrt{1-{g}_{21}^{2}},  \notag \\
\vec{e}_{3} &=&(\vec{g}_{3}-g_{31}\vec{g}_{1}-(\vec{g}_{3}\cdot \vec{e}_{2})%
\vec{g}_{2})/\sqrt{1-g_{31}^{2}-(\vec{g}_{3}\cdot \vec{e}_{2})^{2}},  \notag
\\
\vec{g}_{3}\cdot \vec{e}_{2} &=&(g_{32}-g_{21}g_{31})/\sqrt{1-g_{21}^{2}}.
\label{GS5}
\end{eqnarray}%
where 
\begin{equation}
g_{ik} \equiv \vec{g}_{i}\cdot \vec{g}_{k}=\frac{\gamma _{ik}}{\sqrt{%
\gamma _{ii}\gamma _{kk}}}
\end{equation}
and
\begin{eqnarray}
\gamma _{ik} & \equiv&  \gamma (u_{i},u_{k})=\vec{\Lambda}_{N+1}(u_{i})\cdot 
\vec{\Lambda}_{N+1}(u_{k})  \notag \\
&=&\min(u_{i},u_{k})-u_{i}u_{k}-\sum_{n=1}^{N}\Lambda_{n}(u_{i})\Lambda_{n}(u_{k}). 
 \label{GS6}
\end{eqnarray}%
Using the orthonormal set of constructed vectors, Eq. (\ref{eq:sp}) becomes
\begin{equation}
\vec{a}_{N+1}\cdot \vec{\Lambda}_{N+1}(u_{k})=\sqrt{\varepsilon (u_{k})}%
\sum_{i=1}^{k}\alpha _{ki}\vec{a}_{N+1}\cdot \vec{e}_{i}=\sqrt{\varepsilon
(u_{k})}\sum_{i=1}^{k}\alpha _{ki}\xi _{i}
\end{equation}%
where $\xi _{i}$ denotes a projection of vector $\vec{a}_{N+1}$ onto $\vec{e}%
_{i}$. 

The TI in Eq. (\ref{1.5}) can be geometrically interpreted as an
integration in an infinite-dimensional vector space whose elements are all
possible vectors $\vec{a}_{N+1}$. If we choose $\{\vec{e}_{1},\cdots ,\vec{e}%
_{p}\}$ as the first $p$ basis vectors, then an arbitrary vector will have
the first $p$ components $\{\xi _{1},\cdots ,\xi _{p}\}$. Because the arguments of the potential function do not depend on the
orthogonal projections, the integral with respect to the infinite set
of components $\{\xi _{p+1},\xi _{p+2},\cdots \}$ in the directions
orthogonal to the span of $\{\vec{e}_{1},\cdots ,\vec{e}_{p}\}$ can be
evaluated analytically in Eq. (\ref{eq:ti}) to give unity. As a result the TI in Eq. (\ref{1.5}) is reduced to a $%
p$-dimensional Gaussian transform of the product of $p$ potential
functions%
\begin{eqnarray}
\mu _{p} &=&\overline{G_{p}(u_{1},\ldots ,u_{p})}\equiv
\int\limits_{0}^{1}\cdots \int\limits_{0}^{1}du_{1}\ldots
du_{p}G_{p}(u_{1},\ldots ,u_{p}),  \label{1.6.0} \\
G_{p} &(u_{1},\ldots ,u_{p})\equiv &\left\langle
\prod\limits_{k=1}^{p}V(x_{N}(u_{k})+\sigma \vec{a}_{N+1}\cdot \vec{\Lambda}%
_{N+1}(u_{k}))\right\rangle _{TI}  \notag \\
&=&\prod\limits_{j=1}^{p}\left( \int\limits_{-\infty }^{\infty }\frac{d\xi
_{j}}{\sqrt{2\pi }}\exp \left( -\frac{1}{2}\xi _{j}^{2}\right) \right) \notag \\
&\times&
\prod\limits_{k=1}^{p}V\left[ x_{N}(u_{k})+\sigma \sqrt{\varepsilon (u_{k})}%
\sum_{i=1}^{k}\alpha _{ki}\xi _{i}\right] .  \label{1.6}
\end{eqnarray}%

We can modify the arguments of the potential functions in Eq. (\ref{1.6}) to produce the partial average expression\cite{coalson_partial_1986} for the density matrix derived previously using alternate methods.  We define $\Delta_k$ using the expressions
\begin{equation}
x_{N}(u_{k})+\sigma \sqrt{\varepsilon (u_{k})}\sum_{i=1}^{k}\alpha _{ki}\xi
_{i}=x_{N}(u_{k})+\sigma \sqrt{\varepsilon (u_{k})}\xi _{k}+\Delta _{k},
\label{1.6.1}
\end{equation}%
where%
\begin{equation}
\Delta _{k}=\sigma \sqrt{\varepsilon (u_{k})}\left\{ 
\begin{array}{c}
0,\quad k=1 \\ 
\sum_{i=1}^{k-1}\alpha _{ki}\xi _{i}+(\alpha _{kk}-1)\xi _{k},\quad k\geqq 2%
\end{array}%
\right.  \label{1.6.2}
\end{equation}
From the definition of the coefficients $\alpha _{ki}$, $i=1,\ldots ,k-1$, it is evident the coefficients are small for large $N$.  Then for
$N\rightarrow \infty $, we observe that at $k\geqq 2$ 
\begin{equation}
\Delta _{k}\sim \sigma \sqrt{\varepsilon (u_{k})}\left(
\sum_{i=1}^{k-1}\alpha _{ki}\xi _{i}-\frac{1}{2}\xi
_{k}\sum_{i=1}^{k-1}\alpha _{ki}^{2}\right) . \label{eq:ugh}
\end{equation}%
We can obtain the explicit asymptotic behavior for Eqs. (\ref{GS3}) and (\ref{eq:ugh}) by using the following asymptotic expressions derived in Appendix A.
\begin{equation}
I_{2}(k)\equiv \int\limits_{0}^{1}\int\limits_{0}^{1}du_{1}du_{2}\left[
\gamma (u_{1},u_{2})\right] ^{k}f_{1}(u_{1})f_{2}(u_{2})=\left\{ 
\begin{array}{c}
O\left( \dfrac{1}{N^{k+2}}\right) ,\quad k\;\mathrm{is\;odd} \\ 
O\left( \dfrac{1}{N^{k+1}}\right) ,\quad k\;\mathrm{is\;even}%
\end{array}%
\right.  \label{1.6.5}
\end{equation}%
where $f_{i}(u_{i})$, $i=1,2$ are smooth functions and $k=1,2\ldots $.
From Eqs. (\ref{GS3}), (\ref{eq:ugh}) and asymptotic estimates Eq. (\ref{1.6.5}%
), it follows that%
\begin{eqnarray}
\alpha _{ki} &\sim &g_{ki}=\gamma (u_{k},u_{i})/\sqrt{\varepsilon
(u_{k})\varepsilon (u_{i})},  \notag \\
\gamma _{ki} &\sim &\frac{1}{N^{3}},\quad \gamma _{ki}^{2}\sim \frac{1}{N^{3}%
}  \notag
\end{eqnarray}%
and 
\begin{equation*}
\Delta _{k}\sim \gamma ^{2}(u_{k},u_{i})/\sqrt{\varepsilon
(u_{k})\varepsilon ^{2}(u_{i})}\sim 1/N^{3/2}.
\end{equation*}%
If we neglect the $\Delta _{k}$ terms at $k \ge 2$ in (\ref{1.6}) we find that the
integration over $\xi _{1},\ldots ,\xi _{p}$ variables in Eq. (\ref{1.6}) is
reduced to a product of $p$ one-dimensional integrals so that $\mu _{p}=\mu
_{1}^{p}$. Consequently, the infinite power series is summed to produce an exponential
function%
\begin{equation}
1+\sum_{p=1}^{\infty }\frac{(-\beta )^{p}}{p!}\mu _{1}^{p}=\exp (-\beta \mu
_{1}),  \label{1.6.3}
\end{equation}%
giving the so-called partial averaging (PA) formula first obtained by Doll
{\em et al.}\cite{doll_mboxfourier_1985}

\subsection{The cumulant expansion of the density matrix}

As shown elsewhere\cite{coalson_cumulant_1989} the partial average method expressed in Eq.(\ref{1.6.3}) is the first term in the cumulant expansion of the density matrix.  In this subsection we review the cumulant expansion and derive the asymptotic convergence characteristics of the various cumulant terms for the Fourier representation of the quantum density matrix.  Much of what appears in the initial part of this subsection, in particular the defining relations for the cumulants, can be found elsewhere in the literature.\cite{kubo_generalized_1962}  However, we find it useful to spell out some well-known details to make the subsequent notation and discussion clear.

The cumulant expansion can be obtained from Eq.(\ref{1.4}) by writing
\begin{equation}
\left\langle \exp (-\beta \overline{V})\right\rangle
_{TI}=1+\sum_{p=1}^{\infty }\frac{(-\beta )^{p}}{p!}\mu _{p} \label{eq:cum} \equiv \exp (%
\mathcal{V}_{c})
\end{equation}%
where
\begin{eqnarray}
\mathcal{V}_{c} &=&\ln \left( 1+\sum_{p=1}^{\infty }\frac{(-\beta )^{p}}{p!%
}\mu _{p}\right)  \notag \\
&=&-\sum\limits_{k=1}^{\infty }\frac{1}{k}\left[ -\sum\limits_{p=1}^{\infty }%
\frac{(-\beta )^{p}}{p!}\mu _{p}\right] ^{k}  \notag \\
&=&\sum\limits_{k=1}^{\infty }\frac{(-\beta )^{k}}{k!}\mu _{ck}.  \label{1.7}
\end{eqnarray}%
The last line of Eq. (\ref{1.7}) is obtained by collecting the
terms with the same power $k$ of the potential function [or parameter $\beta 
$] in%
\begin{equation}
\mu _{ck}=\sum\limits_{r=1}^{k}\frac{(-1)^{r+1}}{r}\sum\limits_{p_{1},\cdots
,p_{r}\in C_{kr}}\frac{k!}{p_{1}!\cdots p_{r}!}\mu _{p_{1}}\cdots \mu
_{p_{r}}.  \label{1.7.1}
\end{equation}%
The summation of the integer indices $p_{1},\cdots ,p_{r}\geqslant 1$ is
restricted by the relation%
\begin{equation}
C_{kr}=\left\{ (p_{1},\cdots ,p_{r}):\sum_{j=1}^{r}p_{j}=k\,\right\}
\label{1.7.2}
\end{equation}%
Using Eqs. (\ref{1.7.1}) and (\ref{1.7.2}), the relations between the first four cumulants and the moments are
\begin{equation}
\begin{array}{ccc}
\mu _{c1} & = & \mu _{1}, \\ 
\mu _{c2} & = & \mu _{2}-\mu _{1}^{2}, \\ 
\mu _{c3} & = & \mu _{3}-3\mu _{2}\mu _{1}+2\mu _{1}^{3}. \\ 
\mu _{c4} & = & \mu _{4}-(4\mu _{3}\mu _{1}+3\mu _{2}^{2})+12\mu _{2}\mu
_{1}^{2}-6\mu _{1}^{4}%
\end{array}
\label{1.8}
\end{equation}%
Given that the PA approximation is the first-order cumulant, the errors in the partial average method are determined by the
second, third and higher-order cumulant terms.  In a similar fashion the error in the second-order cumulant is determined by the third, fourth and higher order terms.
Using Eqs. (\ref{1.8}), one can also express the moments in terms of the
cumulants%
\begin{equation}
\begin{array}{ccc}
\mu _{1} & = & \mu _{c1}, \\ 
\mu _{2} & = & \mu _{c2}+\mu _{c1}^{2}, \\ 
\mu _{3} & = & \mu _{c3}+3\mu _{c2}\mu _{c1}+\mu _{c1}^{3}. \\ 
\mu _{4} & = & \mu _{c4}+4\mu _{c3}\mu _{c1}+3\mu _{c2}^{2}+6\mu _{c2}\mu
_{c1}^{2}+\mu _{c1}^{4}%
\end{array}
\label{1.8.1}
\end{equation}

These equations relating the cumulants and the moments can also be interpreted as recurrence relations that allow the
expression of the higher-order cumulants in terms of lower-order ones. A general
expression for the $p^{th}$ order moment can be derived by comparing
coefficients in the power series expansion%
\begin{eqnarray}
1+\sum_{p=1}^{\infty }\frac{x^{p}}{p!}\mu _{p} &=&\exp \left(
\sum\limits_{k=1}^{\infty }\frac{x^{k}}{k!}\mu _{ck}\right)  \notag \\
&=&\sum\limits_{j=0}^{\infty }\frac{1}{j!}\left( \sum\limits_{k=1}^{\infty }%
\frac{x^{k}}{k!}\mu _{ck}\right) ^{j}  \label{1.18.2}
\end{eqnarray}%
From Eq. (\ref{1.18.2}) we obtain%
\begin{eqnarray}
\mu _{p} &=&\frac{d^{p}}{dx^{p}}\sum\limits_{j=1}^{\infty }\frac{1}{j!}%
\left. \left( \sum\limits_{k=1}^{\infty }\frac{x^{k}}{k!}\mu _{ck}\right)
^{j}\right\vert _{x=0}  \notag \\
&=&\frac{1}{1!}\sum\limits_{k_{1}=1}^{\infty }\frac{\mu _{ck_{1}}}{k_{1}!}%
\left. \frac{d^{p}x^{k_{1}}}{dx^{p}}\right\vert _{x=0}+\frac{1}{2!}%
\sum\limits_{k_{1},k_{2}=1}^{\infty }\frac{\mu _{ck_{1}}\mu _{ck_{2}}}{%
k_{1}!k_{2}!}\left. \frac{d^{p}x^{k_{1}+k_{2}}}{dx^{p}}\right\vert
_{x=0}+\ldots  \notag \\
&=&\mu _{cp}+\frac{1}{2!}\sum\limits_{\substack{ k_{1},k_{2}=1  \\ %
k_{1}+k_{2}=p}}\frac{p!}{k_{1}!k_{2}!}\mu _{ck_{1}}\mu _{ck_{2}}+\ldots 
\notag \\
&&+\frac{1}{r!}\sum\limits_{\substack{ k_{1},\ldots ,k_{r}=1  \\ %
k_{1}+\ldots +k_{r}=p}}\frac{p!}{k_{1}!\ldots k_{r}!}\mu _{ck_{1}}\ldots \mu
_{ck_{r}}+\ldots +\mu _{c1}^{p}  \label{1.18.3}
\end{eqnarray}%
The general term in Eq. (\ref{1.18.3}) contains the factor 
\begin{equation*}
\frac{p!}{k_{1}!\ldots k_{r}!},
\end{equation*}%
which defines the number of ways of grouping $p$ objects into $r$ groups of
sizes $k_{1},k_{2},\ldots ,k_{r},$ when the order within each \ group does
not matter. These combinatorics imply that Eq. (\ref%
{1.18.3}) can be represented as a sum over all partitions of $%
\left\{ 1,\ldots ,p\right\} $ vertices

\begin{equation}
\mu _{p}=\sum\limits_{r=1}^{p}\sum\limits_{S_{1},S_{2},\ldots ,S_{r}}\mu
_{c} \left[ S_{1}\right] \mu _{c}\left[ S_{2}\right] \ldots \mu _{c}\left[
S_{r}\right]  \label{1.18.5}
\end{equation}%
where we have introduced a partition $S_{1},S_{2},\ldots ,S_{r}$ of a set of
natural numbers $\left\{ 1,\ldots ,p\right\}.$
% $ such that%
%\begin{equation}
%\begin{array}{l}
%\mathrm{i})\quad \mathrm{each}\quad S_{\alpha }\neq \varnothing \\ 
%\mathrm{ii})\quad \mathrm{each}\quad S_{\alpha }\subset \left\{ 1,\ldots
%,p\right\} \\ 
%\mathrm{iii})\quad S_{\alpha }\bigcap S_{\beta }=\varnothing \quad \mathrm{if%
%}\quad \alpha \neq \beta \\ 
%\mathrm{iv})\quad \bigcup\limits_{\alpha =1,\ldots ,r}S_{\alpha }=\left\{
%1,\ldots ,p\right\}%
%\end{array}
%\label{1.18.4}
%\end{equation}%
%Condition iii) means that partition $S_{1},S_{2},\ldots ,S_{r}$ consist of
%non-overlapping sets.\ Moreover, if a partition $%
As discussed elsewhere,\cite{kubo_generalized_1962} 
if a partition
$S_{1},S_{2},\ldots ,S_{r}$ contains, respectively, $k_{1},k_{2},\ldots
,k_{r} $ elements such that $k_{1}+k_{2}+\ldots +k_{r}=p$, then $\mu _{c}
\left[ S_{1}\right] =\mu _{ck_{1}},\ldots ,\mu _{c}\left[ S_{r}\right] =\mu
_{ck_{r}}$.
The term in Eq. (\ref{1.18.5}) for $r=1$ corresponds to 
$\mu _{cp}$. The resulting expression is
\begin{equation}
\mu _{cp}=\mu _{p}-\sum\limits_{r=2}^{p}\sum\limits_{S_{1},S_{2},\ldots
,S_{r}}\mu _{c}\left[ S_{1}\right] \mu _{c}\left[ S_{2}\right] \ldots \mu
_{c}\left[ S_{r}\right]  \label{1.18.6}
\end{equation}%
where the summation is performed over the ``proper partitions" when $r\geqslant
2$. To illustrate the idea of partitions, we show in Fig. \ref{fig:one} all possible
partitions for case $p=4$ by connecting vertices $1,\ldots ,4$ by lines.  Figure \ref{fig:one} can be compared
to the last line of Eq. (\ref{1.8.1}). The first line from the top in Fig. \ref{fig:one} displays the ``improper partition"  $S_{1}=(1,2,3,4)$ at $r=1$, which corresponds to the term $\mu_{c4}$ in Eq. (\ref{1.8.1}). The partitions $S_{1},S_{2}$ at $r=2$ divide  vertices into two groups. The first group shown in the second line consists of a cluster of three vertices and a cluster containing a single vertex.  The second group shown in the third line consists of two clusters each containing two vertices.  The fourth and fifth lines show the partitions $S_{1},S_{2},S_{3}$ at $r=3$ consisting of a cluster of two vertices and two clusters each containing a single vertex. Finally, in the last line we find the partition $S_{1},S_{2},S_{3},S_{4}$ at $r=4$ consisting of 4 clusters each containing a single vertex.

\subsection{Asymptotic convergence characteristics of the cumulant terms}

A key concern of any numerical path integral approach is the rate of convergence to the exact result with respect to the number of path variables included.  In previous work\cite{eleftheriou_asymptotic_1999-1, predescu_asymptotic_2003} we have examined the Fourier path integral method along with partial averaging and gradient partial averaging, and examined the asymptotic convergence rates with respect to the number of included Fourier coefficients.  As mentioned previously, the partial average method is the first-order term in the cumulant expansion.  The motivation of the current work is to derive the general convergence characteristics of the cumulants truncated at any finite order.  

We ignore the important and interesting question of the convergence of the cumulant expansion itself when a finite number of Fourier coefficients are included.  We make the assumption that if we truncate the cumulant expansion at a finite order, the resulting expression for the quantum density matrix will converge to the exact quantum density matrix in the limit of an infinite set of Fourier coefficients.  

In the following we first examine the convergence of the first three cumulants, and then derive an expression for the asymptotic convergence rates for cumulants of arbitrary, finite order.  The analysis that follows is complex, mainly because the number of terms in high order cumulants is large.  Many of the terms in each order have mixed asymptotic convergence characteristics along with internal cancellations within each order.  We find a diagrammatic approach simplifies the complexity of the problem.

Because 1) we are interested in the asymptotic convergence rate with respect to the number of included path variables $N$, 2) $\varepsilon (u)\sim 1/N$, and 3) the arguments of the
potential functions in Eq. (\ref{1.6}) contain $\varepsilon (u)$, we make
use of the Taylor series expansion for $V$ around the point $x_{N}(u_{k})$. For
simplicity we assume that the potential function $V$ has a convergent Taylor series
to all orders.  As a minimum requirement the potential energy must be infinitely differentiable.

\subsubsection{The first-order cumulant} \label{sec:foc}

Using the Taylor series expansion, we have%
\begin{eqnarray}
\mu _{c1} &=&\int\limits_{0}^{1}du\int\limits_{-\infty }^{\infty }\dfrac{%
d\xi }{\sqrt{2\pi }}\exp \left( -\dfrac{\xi ^{2}}{2}\right) V\left[
x_{N}(u)+\sigma \sqrt{\varepsilon (u)}\xi \right]  \notag \\
&=&\int\limits_{0}^{1}du\sum\limits_{k=0}^{\infty }\frac{V^{(k)}(x_{N}(u))}{%
k!}\sigma ^{k}\varepsilon ^{k/2}(u)\int\limits_{-\infty }^{\infty }\frac{%
d\xi }{\sqrt{2\pi }}\xi ^{k}\exp \left( -\frac{1}{2}\xi ^{2}\right)  \notag
\\
&=&\sum\limits_{k=0}^{\infty }\frac{\sigma ^{2k}}{(2k)!!}\int%
\limits_{0}^{1}duV^{(2k)}(x_{N}(u))\varepsilon ^{k}(u),  \label{1.9}
\end{eqnarray}%
where the first line in Eq.(\ref{1.9}) contains the ``partial averaged potential,''$V_{PA}[x_N(u)]$, which is the Gaussian transform of the ordinary potential
\begin{equation}
V_{PA}[x_N(u)]=\int\limits_{-\infty }^{\infty }\dfrac{%
d\xi }{\sqrt{2\pi }}\exp \left( -\dfrac{\xi ^{2}}{2}\right) V\left[
x_N(u)+\sigma \sqrt{\varepsilon (u)}\xi \right], \label{eq:paap}
\end{equation}
and where we have denoted $V^{(k)}(x)\equiv d^{k}V(x)/dx^{k}$ and $(2k)!!=2\cdot
4\cdot \ldots \cdot 2k=2^{k}k!$ if $k=1,2,\ldots $ [$(2k)!!=1$ at $k=0$].
The Gaussian integral in Eq. (\ref{1.9}) can be evaluated analytically resulting in the expression
\begin{equation}
\int\limits_{-\infty }^{\infty }\frac{d\xi }{\sqrt{2\pi }}\xi ^{k}\exp
\left( -\frac{1}{2}\xi ^{2}\right) =\left\{ 
\begin{array}{c}
0\quad \mathrm{if\;k}\text{\ is\ odd} \\ 
(k-1)!!\quad \mathrm{if\;k}\text{\ is\ even}%
\end{array}%
\right. .  \label{1.10}
\end{equation}%
Here $(2k-1)!!=1\cdot 3\cdot \ldots \cdot (2k-1)$ if $k=1,2,\ldots $ [$%
(2k-1)!!=1$ at $k=0$]. Because integrals of $\varepsilon ^{k}(u)$ multiplied by a
smooth function results in terms of order $1/N^{k}$, the last line in
Eq. (\ref{1.9}) gives the sought asymptotic expansion in powers of $1/N$. 

There is an alternate derivation of the asymptotic expansion that in later developments we find to be particularly convenient for further generalizations. We rewrite the
potential function as 
\begin{equation}
V\left[ x_{N}(u)+\sigma \sqrt{\varepsilon (u)}\xi \right] =\exp \left( \sqrt{%
\varepsilon (u)}\xi \hat{d}\right) V(x_{N}(u))  \label{1.10.0}
\end{equation}%
where $\displaystyle \hat{d}\equiv \sigma \frac{d}{dx}$ is a differential operator. Because $\hat{d}$-operators generate a commutative algebra similar to
that of $c$-numbers, the resulting
operator integral over $\xi $ can be calculated as an ordinary Gaussian
integral, 
\begin{eqnarray}
g_{1}(u,\hat{d}) &\equiv &\int\limits_{-\infty }^{\infty }\dfrac{d\xi }{%
\sqrt{2\pi }}\exp \left( -\dfrac{\xi ^{2}}{2}+\sqrt{\varepsilon (u)}\xi \hat{%
d}\right)  \notag \\
&=&\exp \left( \dfrac{\varepsilon (u)}{2}\hat{d}^{2}\right) .  \label{1.10.1}
\end{eqnarray}%
Applying the right-hand-side of Eq. (\ref{1.10.1}) to the potential
function, we obtain%
\begin{equation}
\mu _{c1}=\int\limits_{0}^{1}dug_{1}(u,\hat{d})V(x_{N}(u))=\int%
\limits_{0}^{1}du\exp \left( \dfrac{\varepsilon (u)}{2}\hat{d}^{2}\right)
V(x_{N}(u)).  \label{1.10.2}
\end{equation}%
Using a Taylor series to expand the exponential, the result is the
asymptotic expansion for the PA potential
\begin{eqnarray}
V_{PA}(x_{N}(u)) &=&\exp \left( \dfrac{\varepsilon (u)}{2}\hat{d}^{2}\right)
V(x_{N}(u))  \notag \\
&=&\sum\limits_{k=0}^{\infty }\frac{\left( \dfrac{\varepsilon (u)}{2}\hat{d}%
^{2}\right) ^{k}}{k!}V(x_{N}(u)),  \label{1.10.3}
\end{eqnarray}%
formally identical to Eq. (\ref{1.9}).  The primitive Fourier path integral method corresponds to the first term, $k=0$, while the ``gradient partial average'' approximation\cite{doll_equilibrium_1990-1,coalson_partial_1986} is given by the first two terms ($k=0,1$) in the expansion of Eq.(\ref{1.10.3}).

Numerical evaluation of the one-dimensional $u$-integrals can employ any convenient quadrature rule.  We have found the Gauss-Legendre
quadrature formula \cite{press_numerical_1992} 
to be generically useful 
\begin{eqnarray}
\mu_{c1}=\overline{V_{PA}(x_{N}(u))}%
=\sum_{i=1}^{N_{q}}w_iV_{PA}(x_{N}(u_{i}))  \label{1.10.4}
\end{eqnarray}
with $w_i$ being a Gauss-Legendre weight and $N_{q}$ the
number of quadrature points. The evaluation of the integral Eq. (%
\ref{1.10.4}) requires $N_{q}$ calls to calculate the potential function $%
V_{PA}$, whereas the $w_i$ coefficients do not depend on $V_{PA}$ and
can be precalculated.

\subsubsection{The second-order cumulant}

The second-order cumulant includes both the average potential and the average of the square of the potential.  The average potential has already been discussed in Section \ref{sec:foc}, so we consider the average of the square of the potential function $\bar{%
V}$, Eq. (\ref{1.6}) at $p=2$. Using the operator
representation (\ref{1.10.0}) twice, we obtain 
\begin{eqnarray}
G_{2}(u_{1},u_{2}) &=&\left\langle V\left[ x_{N}(u_{1})+\sigma \sqrt{%
\varepsilon (u_{1})}\xi _{1}\right] V\left[ x_{N}(u_{2})+\sigma \sqrt{%
\varepsilon (u_{2})}(\alpha _{21}\xi _{1}+\alpha _{22}\xi _{2})\right]
\right\rangle _{TI}  \notag \\
&=&g_{2}(u_{1},\hat{d}_{1},u_{2},\hat{d}_{2})V(1)V(2)  \label{1.11}
\end{eqnarray}%
\ where%
\begin{eqnarray}
g_{2}(u_{1},\hat{d}_{1},u_{2},\hat{d}_{2}) &=&\int\limits_{-\infty }^{\infty
}\int\limits_{-\infty }^{\infty }\frac{d\xi _{1}d\xi _{2}}{2\pi }\exp \left(
-\frac{\xi _{1}^{2}+\xi _{2}^{2}}{2}\right)  \notag \\
&&\times \exp \left( \sqrt{\varepsilon _{1}}\xi _{1}\hat{d}_{1}\right) \exp
\left( \sqrt{\varepsilon _{2}}(\alpha _{21}\xi _{1}+\alpha _{22}\xi _{2})%
\hat{d}_{2}\right) .  \label{1.12}
\end{eqnarray}% 
In Eqs. (\ref{1.11}) and (\ref{1.12}) we have introduced the notation 
\begin{equation}
V(i)\equiv V(x_{N}(u_{i})),\quad \varepsilon _{i}\equiv \varepsilon
(u_{i}),\quad \hat{d}_{i}=\sigma \frac{d}{dx_{i}},\quad i=1,2  \label{1.11.1}
\end{equation}%
The differential operators $\hat{d}_{1}$ and $\hat{d}_{2}$ act,
correspondingly, on potential functions $V(1)$ and $V(2)$. The operators $\hat{d%
}_{1}$ and $\hat{d}_{2}$ commute and generate a
commutative operator algebra. Evaluating the Gaussian integral Eq. (\ref%
{1.12}), we obtain%
\begin{eqnarray}
g_{2}(u_{1},\hat{d}_{1},u_{2},\hat{d}_{2}) &=&\exp \left( \dfrac{1}{2}\left[
\varepsilon _{1}\hat{d}_{1}^{2}+\varepsilon _{2}\hat{d}_{2}^{2}\right]
+\gamma _{21}\hat{d}_{1}\hat{d}_{2}\right)  \notag \\
&=&\exp \left( \dfrac{1}{2}\sum\limits_{i,j=1}^{2}\gamma _{ij}\hat{d}_{i}%
\hat{d}_{j}\right)  \label{1.13}
\end{eqnarray}%
where $\gamma _{21}$ is defined in Eq. (\ref{GS6}). In the second line,
we have made use of the symmetry property $\gamma _{ij}=\gamma _{ji}$ and the relation $%
\varepsilon _{i}=\varepsilon (u_{i})=\gamma (u_{i},u_{i})=\gamma _{ii}$. \
Using the defining relation expressed in Eq. (\ref{1.10.1}), Eq. (\ref{1.13}) can also be rewritten as%
\begin{equation}
g_{2}(u_{1},\hat{d}_{1},u_{2},\hat{d}_{2})=g_{1}(u_{1},\hat{d}%
_{1})g_{1}(u_{2},\hat{d}_{2})f_{2}(\Gamma _{21})
\end{equation}%
where%
\begin{eqnarray}
f_{2}(\Gamma _{21}) &=&\exp \left( \Gamma _{21}\right) , \\
\Gamma _{21} &=&\gamma _{21}\hat{d}_{1}\hat{d}_{2}.
\end{eqnarray}
From Eqs. (\ref{1.6.0}), (\ref{1.10.2}) and (\ref{1.11}) we have%
\begin{eqnarray}
\mu _{2} &=&\overline{G_{2}(u_{1},u_{2})},  \notag \\
G_{2}(u_{1},u_{2}) &=&g_{1}(u_{1},\hat{d}_{1})g_{1}(u_{2},\hat{d}%
_{2})f_{2}(\Gamma _{21})V(1)V(2)  \notag \\
&=&f_{2}(\Gamma _{21})V_{PA}(1)V_{PA}(2),  \label{1.14} \\
\mu _{1}^{2}
&=&\int\limits_{0}^{1}\int\limits_{0}^{1}du_{1}du_{2}g_{1}(u_{1},\hat{d}%
_{1})g_{1}(u_{2},\hat{d}_{2})V(1)V(2)  \notag \\
&=&\int\limits_{0}^{1}\int\limits_{0}^{1}du_{1}du_{2}V_{PA}(1)V_{PA}(2)
\label{1.15}
\end{eqnarray}%
and%
\begin{eqnarray}
\mu _{c2} &=&\mu _{2}-\mu _{1}^{2}=\overline{G_{c2}(u_{1},u_{2})}  \notag \\
G_{c2}(u_{1},u_{2}) &=&f_{c2}(\Gamma _{21})V_{PA}(1)V_{PA}(2)  \label{1.16}
\end{eqnarray}%
where%
\begin{equation}
f_{c2}(\Gamma _{21})=f_{2}(\Gamma _{21})-1=\exp \left( \Gamma _{21}\right) -1.
\label{1.17}
\end{equation}%
The integrand in $\mu _{1}^{2}$ is separable in the $u_{1}$ and $%
u_{2}$-time variables%
\begin{equation}
V_{PA}(x_{N}(u_{1}))V_{PA}(x_{N}(u_{2}))\equiv V_{PA}(1)V_{PA}(2).
\label{1.17.1}
\end{equation}% 
The integrations with respect to the variables $\mu _{1}^{2}$ over $u_{1}$ and $u_{2}$ $\ $%
are independent, and we call such terms that factor
{\em uncorrelated}. In contrast $\mu _{c2}$ is
proportional to $\gamma _{21}$%
\begin{eqnarray}
G_{c2}(u_{1},u_{2}) &=&\left[ \exp \left( \Gamma _{21}\right) -1\right]
V_{PA}(1)V_{PA}(2)  \notag \\
&=&\sum\limits_{k=1}^{\infty }\frac{\Gamma _{21}^{k}}{k!}V_{PA}(1)V_{PA}(2) 
\notag \\
&=&\sum\limits_{k=1}^{\infty }\frac{\gamma _{21}^{k}}{k!}\left[ \hat{d}%
_{1}^{k}V_{PA}(1)\right] \left[ \hat{d}_{2}^{k}V_{PA}(2)\right],
\label{1.17.2}
\end{eqnarray}%
and
\begin{eqnarray}
\gamma _{21} &=&2\sum\limits_{n=N+1}^{\infty }\frac{\sin (\pi nu_{1})\sin
(\pi nu_{2})}{(\pi n)^{2}}  \notag \\
&=&\sum\limits_{n=N+1}^{\infty }\frac{\cos \pi n(u_{2}-u_{1})-\cos \pi
n(u_{2}+u_{1})}{(\pi n)^{2}}  \label{1.18}
\end{eqnarray}%
cannot be factored into a product of two separate functions of the time
variables $u_{1}$ and $u_{2}$. The function $\gamma_{21}$ establishes a time
correlation between time variables $u_{1}$ and $u_{2}$. The integrand Eq. (%
\ref{1.17.2}) does not contain uncorrelated terms. All the uncorrelated
terms cancel; {\em i.e.}  if $\gamma _{21}\to 0$, then $%
G_{c2}(u_{1},u_{2})\to 0$ and $\mu _{c2}\to 0$.

The demonstrated dependence of $\mu_{c2}$ on the operator $\Gamma_{21}$ can
be graphically illustrated using a diagrammatic notation that proves to be especially valuable in simplifying the algebra for the higher-order cumulants.  For $\mu_{c2}$ the diagrams are shown in Fig. \ref{fig:two} (a). We let small solid circles represent
vertices 1 and 2 with the corresponding vertex functions $V_{PA}(1)$ and $%
V_{PA}(2)$.   The line that connects vertices 1 and 2 in the middle diagram of Fig. \ref{fig:two} (a) represents the interaction 
via $\Gamma_{21}$, which acts on the vertex functions via the differential operators $\hat{d}_1$ and $\hat{d}_2$.   The two lines connecting vertices 1 and 2  shown in the right diagram, correspond to $\Gamma_{21}^2$ entering in Eq. (\ref{1.17.2}) with the weight coefficient $1/2!$. In general, the $k^{th}$ power of a $\Gamma_{12}$ term, with weight coefficients $1/k!$, can be conveniently depicted by a diagram with the $k$ lines connecting the two vertices. Summing the diagrams with the number of connecting lines from $k=1$ up  to $\infty$  produces a diagrammatic representation of the series expansion given in Eq. (\ref{1.17.2}).
Corresponding to these two-vertex connected diagrams, we define a second-order
cumulant function $\mu_{c2}(\Gamma_{21})$, with the characteristic property that 
$\mu_{c2}(\Gamma_{21})=0$ if $\Gamma_{21}= 0$.  In other words, the cumulant
function takes a zero value on the corresponding disconnected vertex
diagram, which is defined as two vertices not connected by a line
that corresponds to the limiting case $\Gamma_{21}\to 0$ [the left diagram in Fig. \ref{fig:two} (a)].

Using the expansion given in Eq. (\ref{1.17.2}), the second cumulant function can be written
as a sum%
\begin{equation}
\mu _{c2}=\sum\limits_{k=1}^{\infty }\mu _{c2}^{(k)}  \label{1.18.1}
\end{equation}
where $\mu _{c2}^{(k)}$ is the corresponding contribution from the $k^{th}$
order derivative of the potential function $V_{PA}$%
\begin{equation}
\mu _{c2}^{(k)}=\int\limits_{0}^{1}\int\limits_{0}^{1}du_{1}du_{2}\frac{%
\gamma _{21}^{k}}{k!}\left[ \hat{d}_{1}^{k}V_{PA}(1)\right] \left[ \hat{d}%
_{2}^{k}V_{PA}(2)\right]  \label{1.18.7}
\end{equation}
Expressions for the integrals given in Eq. (\ref{1.18.7}) are derived in Appendix A [see Eq. (\ref{A1})].  Using the results of Appendix A, we find
\begin{eqnarray}
\mu _{c2}^{(k)} &=&\frac{2^{k}\sigma ^{2k}}{k!}\sum_{n_{1},\cdots
,n_{k}=N+1}^{\infty }\dfrac{1}{(\pi n_{1})^{2}\cdots (\pi n_{k})^{2}}  \notag
\\
&&\times \left[ \int\limits_{0}^{1}du\sin (\pi n_{1}u)\cdots \sin (\pi
n_{k}u)\frac{d^{k}V_{PA}(x_{N}(u))}{dx^{k}}\right] ^{2}  \label{1.18.8}
\end{eqnarray}
from which it follows that each term $\mu _{c2}^{(k)}\geqslant 0$ and, as a
consequence, their sum, Eq. (\ref{1.18.1}), is non-negative.

As discussed for the first-order cumulant [see Eq. (\ref{1.10.4})], $\mu _{c2}^{(k)}$ can be calculated
numerically using the Gauss-Legendre quadrature formula 
\begin{equation}
\mu _{c2}^{(k)}=\sum_{i,j=1}^{N_{q}}\varpi _{ij}^{(k)}\dfrac{%
d^{k}V_{PA}(x_{N}(u_{i}))}{dx^{k}}\dfrac{d^{k}V_{PA}(x_{N}(u_{j}))}{dx^{k}}
\label{1.19.1}
\end{equation}%
where the symmetric semi-positive definite matrix 
\begin{equation}
\varpi _{ij}^{(k)}\equiv \frac{\sigma ^{2k}}{k!}w_i\gamma
^{k}(u_{i},u_{j})w_j  \label{1.19.2}
\end{equation}
does not depend on the potential function, and its matrix elements can be
precalculated. The evaluation of the double integral sum in Eq. (\ref%
{1.19.1}) requires $N_{q}$ calls to calculate the $k^{th}$ order derivative of
the potential function $V_{PA}$. Using the Cholesky decomposition\cite{golub_matrix_1996}
\begin{equation}
\varpi ^{(k)}=\eta ^{(k)}\eta ^{(k)T}  \label{1.19.3}
\end{equation}
where $\eta ^{(k)}$ is a unique lower triangular matrix with positive
diagonal entries, Eq. (\ref{1.19.1}) can be rewritten as 
\begin{equation}
\mu _{c2}^{(k)}=\sum_{i=1}^{N_{q}}\left[\sum_{j=1}^{i} \eta ^{(k)}_{ji} 
\dfrac{d^{k}V_{PA}(x_{N}(u_{j}))}{dx^{k}}\right] ^{2}  \label{1.19.4}
\end{equation}

From Eq. (\ref{1.6.5}) follows that $\mu _{c2}^{(1)}$ and $\mu
_{c2}^{(2)}$ are of the same order of magnitude, $1/N^{3}$, and 
\begin{equation}
\mu _{c2}=\mu _{c2}^{(1)}+\mu _{c2}^{(2)}+O\left( \frac{1}{N^{5}}\right) \label{1.19.5}
\end{equation}%
Using the asymptotic estimates, Eqs. (\ref{A8})-(\ref{A8.4}) and (\ref{A9})-(\ref%
{A13}), derived in Appendix A, the asymptotically leading terms $\mu
_{c2}^{(1)}$ and $\mu _{c2}^{(2)}$, can be written as%
\begin{eqnarray}
\mu _{c2}^{(1)} &=&\sigma
^{2}\int\limits_{0}^{1}\int\limits_{0}^{1}du_{1}du_{2}\gamma _{21}\dfrac{%
dV_{PA}(x_{N}(u_{1}))}{dx}\dfrac{dV_{PA}(x_{N}(u_{2}))}{dx}  \notag \\
&=&\sum\limits_{n=N+1}^{\infty }\frac{2\sigma ^{2}}{(\pi n)^{4}}\left[ \frac{%
dV(x)}{dx}-(-1)^{n}\frac{dV(x^{\prime })}{dx}\right] ^{2}+O\left( \frac{1}{%
N^{5}}\right)  \notag \\
&=&\frac{2\sigma ^{2}}{3\pi ^{4}N^{3}}\left[ \left( \frac{dV(x)}{dx}\right)
^{2}+\left( \frac{dV(x^{\prime })}{dx}\right) ^{2}\right]  \notag \\
&+&\dfrac{2(-1)^{N}\sigma ^{2}}{\pi ^{4}N^{4}}\frac{dV(x)}{dx}\frac{%
dV(x^{\prime })}{dx}+O\left( \frac{1}{N^{5}}\right) ,  \label{1.21}
\end{eqnarray}%
and
\begin{eqnarray}
\mu _{c2}^{(2)} &=&\frac{\sigma ^{4}}{2}\int\limits_{0}^{1}\int%
\limits_{0}^{1}du_{1}du_{2}\gamma _{21}^{2}\dfrac{d^{2}V_{PA}(x_{N}(u_{1}))}{%
dx^{2}}\dfrac{d^{2}V_{PA}(x_{N}(u_{2}))}{dx^{2}}  \notag \\
&=&\dfrac{\sigma ^{4}}{6\pi ^{4}N^{3}}\int\limits_{0}^{1}du\left[ \dfrac{%
d^{2}V_{PA}(x_{N}(u))}{dx^{2}}\right] ^{2}+O\left( \frac{1}{N^{5}}\right) ,
\label{1.21.1}
\end{eqnarray}
Combining the above asymptotic formulas, we obtain 
\begin{eqnarray}
\mu _{c2}&=&\frac{\sigma ^{2}}{3\pi ^{4}N^{3}}\left\{ 2\left[ \left( \frac{%
dV(x)}{dx}\right) ^{2}+\left( \frac{dV(x^{\prime })}{dx}\right) ^{2}\right] +%
\frac{\sigma ^{2}}{2}\int\limits_{0}^{1}du\left[ \dfrac{d^{2}V_{PA}(x_{N}(u))%
}{dx^{2}}\right] ^{2}\right\}  \notag \\
&&+\frac{2(-1)^{N}\sigma ^{2}}{\pi ^{4}N^{4}}\frac{dV(x)}{dx}\frac{%
dV(x^{\prime })}{dx}+O\left( \frac{1}{N^{5}}\right) .  \label{1.22}
\end{eqnarray}%
In the first line, we have collected terms of third order, while in the
second line the terms of fourth and higher orders are included. Equation (\ref{1.22}) confirms previous work\cite{predescu_asymptotic_2003} that shows that the
PA approximation has a third-order convergence rate. Our numerical
investigation of the PA convergence rate in a 1$D$ model\cite{kunikeev_numerical_2009}
is consistent with this asymptotic convergence rate.   The calculation of the convergence
constants in Eq. (\ref{1.22}) (the expression in the curly brackets) can be
reduced to a one-dimensional integration over $u$. 

Equation (\ref{1.19.1}) is comprised of $N_q^2$ terms composed of a product of $k^{th}$-order derivatives of the potential energy.  The product of the two derivatives of the same order implies that for each $k$, the $k^{th}$-order derivative of $V_{PA}[x_N(u_i)]$ at each grid point $u_i$ needs to be evaluated only once.  In that way the $N_q^2$ operations required to evaluate $\mu_{c2}^{(k)}$ in Eq. (\ref{1.19.1}) require only $N_q$ evaluations of the derivatives of the potential energy.  Because the computational work can be expected to be dominated by the evaluation of the potential energy and its derivatives, the work required to determine Eqs. (\ref{1.19.1}) and (\ref{1.19.4}) should scale nearly linearly in $N_q$.  From Eq. (\ref{1.19.5}) only $\mu_{c2}^{(1)}$ and $\mu_{c2}^{(2)}$ are required to attain an $N^{-5}$ order asymptotic convergence rate further emphasizing that we can attain the same asymptotic convergence rate as the full cumulant with truncated algorithms that scale nearly linearly in $N_q$.

The above conclusions about the convergence rate are implicitly based on
the assumption that the higher order cumulant terms, $\mu _{ck}$, $%
k\geqslant 3$, contribute at a faster rate. We check this assumption
for the third cumulant term in the next section where we
demonstrate that $\mu _{c3}$ is fifth order in $1/N$.

\subsubsection{The third-order cumulant}

Using methods identical to those employed in the derivation of Eq. (\ref{1.14}), we obtain for the third-order moment
\begin{eqnarray}
\mu _{3} &=&\overline{G_{3}(u_{1},u_{2},u_{3})}  \label{1.23} \\
&=&\int\limits_{0}^{1}\int\limits_{0}^{1}\int%
\limits_{0}^{1}du_{1}du_{2}du_{3}g_{3}(u_{1},\hat{d}_{1},u_{2},\hat{d}%
_{2},u_{3},\hat{d}_{3})V(1)V(2)V(3),
\end{eqnarray}%
where%
\begin{eqnarray}
g_{3} &=&\int\limits_{-\infty }^{\infty }\int\limits_{-\infty }^{\infty
}\int\limits_{-\infty }^{\infty }\frac{d\xi _{1}d\xi _{2}d\xi _{3}}{(2\pi
)^{3/2}}\exp \left( -\frac{1}{2}(\xi _{1}^{2}+\xi _{2}^{2}+\xi _{3}^{2})+%
\sqrt{\varepsilon _{1}}\xi _{1}\hat{d}_{1}\right.  \notag \\
&&+\left. \sqrt{\varepsilon _{2}}(\alpha _{21}\xi _{1}+\alpha _{22}\xi _{2})%
\hat{d}_{2}+\sqrt{\varepsilon _{3}}(\alpha _{31}\xi _{1}+\alpha _{32}\xi
_{2}+\alpha _{33}\xi _{3})\hat{d}_{3}\right) .  \label{1.23.1}
\end{eqnarray}%
Evaluating the Gaussian integral Eq. (\ref{1.23.1}), we obtain%
\begin{eqnarray}
g_{3} &=&\exp \left( \frac{1}{2}\sum\limits_{i=1}^{3}\varepsilon _{i}\hat{d}%
_{i}^{2}+\sum\limits_{i<j=1}^{3}\gamma _{ij}\hat{d}_{i}\hat{d}_{j}\right)
=\exp \left( \frac{1}{2}\sum\limits_{i,j=1}^{3}\gamma _{ij}\hat{d}_{i}\hat{d}%
_{j}\right)  \notag \\
&=&\left[ \prod\limits_{i=1}^{3}g_{1}(u_{i},\hat{d}_{i})\right]
f_{3}(\Gamma_{12},\Gamma_{23},\Gamma_{13})  \label{1.24}
\end{eqnarray}%
where 
\begin{equation}
f_{3}(\Gamma_{12},\Gamma_{23},\Gamma_{13})=\exp
\left(\sum\limits_{i<j=1}^{3}\Gamma _{ij}\right)
\end{equation}
and
\begin{equation}
\Gamma _{ij}=\Gamma _{ji}= \gamma _{ij}\hat{d}_{i}\hat{d}_{j}.
\label{1.24.1}
\end{equation}
Using Eqs. (\ref{1.10.1}) and (\ref{1.24}), we then obtain%
\begin{eqnarray}
G_{3}(u_{1},u_{2},u_{3})=f_{3}(\Gamma_{12},\Gamma_{23},%
\Gamma_{13})V_{PA}(1)V_{PA}(2)V_{PA}(3)
\end{eqnarray}
and 
\begin{eqnarray}
\mu_{c3}&=&\mu_{3}-3\mu_{2}\mu_{1}+2\mu_{1}^{3}=\overline{%
G_{c3}(u_{1},u_{2},u_{3})},  \notag \\
G_{c3}(u_{1},u_{2},u_{3})&=&f_{c3}(\Gamma_{12},\Gamma_{23},%
\Gamma_{13})V_{PA}(1)V_{PA}(2)V_{PA}(3)  \label{1.24.2}
\end{eqnarray}
where 
\begin{eqnarray}
f_{c3}=f_{3}(\Gamma_{12},\Gamma_{23},\Gamma_{13})-f_{2}(\Gamma_{12})-f_{2}(%
\Gamma_{23})-f_{2}(\Gamma_{13})+2.  \label{1.24.3}
\end{eqnarray}

We now show that the uncorrelated terms, $-3\mu _{2}\mu _{1}+2\mu
_{1}^{3}$ sum to zero
in the expression for $\mu _{c3}$.  We introduce the convenient notation%
\begin{equation}
\Gamma _{1}\equiv \Gamma _{23},\quad \Gamma _{2}\equiv \Gamma _{13},\quad
\Gamma _{3}\equiv \Gamma _{12}.  \label{1.28}
\end{equation}%
and rewrite $f_{c3}$%
\begin{eqnarray}
f_{c3}(\Gamma_{1},\Gamma_{2},\Gamma_{3}) &=&\exp \left(
\sum\limits_{i=1}^{3}\Gamma _{i}\right) -\sum\limits_{i=1}^{3}\exp \left(
\Gamma _{i}\right) +2  \notag \\
&=&\sum\limits_{k=1}^{\infty }\frac{\left( \Gamma _{1}+\Gamma _{2}+\Gamma
_{3}\right) ^{k}-\Gamma _{1}^{k}-\Gamma _{2}^{k}-\Gamma _{3}^{k}}{k!}  \notag
\\
&=&\sum\limits_{k=2}^{\infty }\left( \sum\limits_{\substack{ %
p_{1},p_{2},p_{3}\geqslant 0  \\ p_{1}+p_{2}+p_{3}=k}}\frac{\Gamma
_{1}^{p_{1}}\Gamma _{2}^{p_{2}}\Gamma _{3}^{p_{3}}}{p_{1}!p_{2}!p_{3}!}-%
\frac{\sum_{i=1}^{3}\Gamma _{i}^{k}}{k!}\right) .  \label{1.29}
\end{eqnarray}%
The inner summation in Eq. (\ref{1.29}) is restricted to non-negative integer indices 
$p_{1},p_{2},p_{3}$ with the additional requirement $p_{1}+p_{2}+p_{3}=k$. It is clear
from the second line of Eq. (\ref{1.29}) that the term at $k=1$ is zero. If
any two of three indices take a zero value, say, $p_{1}=p_{2}=0$, then $%
p_{3}=k$ and the corresponding term%
\begin{equation}
\left. \frac{\Gamma _{1}^{p_{1}}\Gamma _{2}^{p_{2}}\Gamma _{3}^{p_{3}}}{%
p_{1}!p_{2}!p_{3}!}\right\vert _{(p_{1}=0,p_{2}=0,p_{3}=k)}=\frac{\Gamma
_{3}^{k}}{k!}  \label{1.30}
\end{equation}%
cancels in Eq. (\ref{1.29}). We then find%
\begin{equation}
f_{c3}=\sum\limits_{k=2}^{\infty }\sum\limits_{p_{1},p_{2},p_{3}\in C_{k}}%
\frac{\Gamma _{1}^{p_{1}}\Gamma _{2}^{p_{2}}\Gamma _{3}^{p_{3}}}{%
p_{1}!p_{2}!p_{3}!}  \label{1.31}
\end{equation}%
where the following constraints on the indices are imposed%
\begin{equation}
C_{k}=(p_{1},p_{2},p_{3})=\left\{ 
\begin{array}{c}
p_{1},p_{2},p_{3}\geqslant 0,\quad p_{1}+p_{2}+p_{3}=k, \\ 
\text{at least two of indices are non-zero.}%
\end{array}%
\right.  \label{1.32}
\end{equation}
As a consequence of the constraints, we find Eq. (\ref{1.31}) is proportional to the
cross terms only and does not contain terms arising from the uncorrelated terms, $%
-3\mu _{2}\mu _{1}+2\mu _{1}^{3}$, the terms with at least two zero indices.

As with the second-order cumulant it is useful to introduce a diagrammatic notation for the various contributions. 
We define a diagram with three vertices 1, 2, 3 and corresponding
vertex functions.  In Fig. 2 (b), the three vertices are shown to be connected by lines that correspond
to the pairwise vertex couplings $\Gamma_{12}$, $\Gamma_{23}$, $\Gamma_{31}$.  
As follows from Eq. (\ref{1.24.3}) or (\ref{1.31}) the third
cumulant function $\mu_{c3}(\Gamma_{12},\Gamma_{23},\Gamma_{31})$  is zero whenever the underlying diagram is disconnected. We find disconnected diagrams
when any two of the vertex couplings are zero.  For example, we obtain zero if
$\Gamma_{12}=\Gamma_{23}= 0$, from which it follows that $\mu_{c3}(0,0,\Gamma_{31})=0$.  In
this case vertex 2 is disconnected from vertices 1 and 3, which are
connected by a line that corresponds to the $\Gamma_{31}\ne 0$ coupling.

We now examine the cross terms up to third order; i.e. the terms with $k=2,3$ in Eq. (%
\ref{1.31}).  We find
\begin{eqnarray}
f_{c3} &\approx &\sum\limits_{k=2}^{3}\sum\limits_{p_{1},p_{2},p_{3}\in
C_{k}}\frac{\Gamma _{1}^{p_{1}}\Gamma _{2}^{p_{2}}\Gamma _{3}^{p_{3}}}{%
p_{1}!p_{2}!p_{3}!}  \notag \\
&=&\Gamma _{1}\Gamma _{2}+\Gamma _{2}\Gamma _{3}+\Gamma _{1}\Gamma
_{3}+\Gamma _{1}\Gamma _{2}\Gamma _{3}  \notag \\
&&+\frac{1}{2}\left( \Gamma _{1}^{2}\Gamma _{2}+\Gamma _{1}\Gamma
_{2}^{2}+\Gamma _{2}^{2}\Gamma _{3}+\Gamma _{2}\Gamma _{3}^{2}+\Gamma
_{1}^{2}\Gamma _{3}+\Gamma _{1}\Gamma _{3}^{2}\right) .  \label{1.33}
\end{eqnarray}
 In Fig. \ref{fig:two} (b), we display the diagrams that correspond to the terms in the second line of Eq. (\ref{1.33}).
Using the asymptotic formulas derived in Appendix A [Eqs. (\ref{A22}) and (\ref{A23})],
we find that terms in the second line of Eq. (\ref{1.33}) are fifth order
\begin{eqnarray}
&&\Gamma _{1}\Gamma _{2}+\Gamma _{2}\Gamma _{3}+\Gamma _{1}\Gamma
_{3}+\Gamma _{1}\Gamma _{2}\Gamma _{3}  \notag \\
&=&\gamma _{23}\gamma _{13}\hat{d}_{1}\hat{d}_{2}\hat{d}_{3}^{2}+\gamma
_{13}\gamma _{12}\hat{d}_{1}^{2}\hat{d}_{2}\hat{d}_{3}+\gamma _{23}\gamma
_{12}\hat{d}_{1}\hat{d}_{2}^{2}\hat{d}_{3}  \notag \\
&&+\gamma _{12}\gamma _{23}\gamma _{13}\hat{d}_{1}^{2}\hat{d}_{2}^{2}\hat{d}%
_{3}^{2},  \label{1.34}
\end{eqnarray}%
and the terms in the third line are of the higher order; namely, they produce a sixth order contribution.  For example, if we
examine a typical term of the third line in Eq. (\ref{1.33})
\begin{equation}
\Gamma _{1}^{2}\Gamma _{2}=\gamma _{23}^{2}\gamma _{13}\hat{d}_{1}\hat{d}%
_{2}^{2}\hat{d}_{3}^{3}.
\end{equation}
and considering Eqs. (\ref{1.6.5}), we find that 
\begin{eqnarray}
\gamma _{23}^{2}\gamma _{13}\sim 1/N^{6}
\end{eqnarray}%
and%
\begin{equation}
\Gamma _{1}^{2}\Gamma _{2}\sim \frac{1}{N^{6}}.
\end{equation}
After substituting Eq. (\ref{1.34}) into (\ref{1.33}) and (\ref{1.24.2}),
we find
\begin{eqnarray}
\mu _{c3} &=&\sigma ^{6}\int\limits_{0}^{1} \int\limits_{0}^{1}
\int\limits_{0}^{1} du_{1}du_{2}du_{3}\, \gamma _{12}\gamma _{13}\gamma
_{23} \dfrac{d^2V_{PA}(1)}{dx^2}\dfrac{d^2V_{PA}(2)}{dx^2}\dfrac{d^2V_{PA}(3)%
}{dx^2}  \notag \\
&&+3\sigma^{4} \int\limits_{0}^{1}\int\limits_{0}^{1}\int\limits_{0}^{1}
du_{1}du_{2}du_{3}\,\gamma _{12}\gamma _{13} \dfrac{d^2V_{PA}(1)}{dx^2} 
\dfrac{dV_{PA}(2)}{dx}\ \dfrac{dV_{PA}(3)}{dx}+O\left( \frac{1}{N^{6}}\right)
\label{1.43}
\end{eqnarray}%
In Eq.(\ref{1.43}), the first integral is a contribution from the loop diagram, the second line of Fig. \ref{fig:two} (b), corresponding to the $\Gamma_{12}\Gamma_{23}\Gamma_{31}$ term, whereas the second one represents a combined contribution, which is included by a factor 3 in front of the integral, from the degenerate diagrams shown in the first line of Fig. \ref{fig:two} (b).

As with the second-order cumulant terms, Eq. (\ref{1.43}) implies that we can attain a $N^{-6}$ asymptotic convergence rate by retaining only the first two terms on the right-hand side.  The work required when the integrals are evaluated numerically using Gauss-Legendre quadrature with $N_q$ quadrature points for a single one-dimensional imaginary-time integration, results in a
total of $2N_q$ calls of the first and second derivatives of the
potential function $V_{PA}$. From the asymptotic estimates Eqs. (%
\ref{A22}), (\ref{A23}) obtained in Appendix A for the above fifth-order contributions, the
integrals expressed in Eq. (\ref{1.43}) can be rewritten as%
\begin{eqnarray}
\mu _{c3} &=&\frac{\sigma ^{4}}{5\pi ^{6}N^{5}}\left\{ \sigma
^{2}\int\limits_{0}^{1}du\left[ \dfrac{d^{2}V(x_{N}(u))}{dx^{2}}\right]
^{3}+3\dfrac{d^{2}V(x)}{dx^{2}}\left[ \dfrac{dV(x)}{dx}\right] ^{2}+3\dfrac{%
d^{2}V(x^{\prime })}{dx^{2}}\left[\dfrac{dV(x^{\prime})}{dx} \right]
^{2}\right\}  \notag \\
&&+O\left( \frac{1}{N^{6}}\right) .  \label{1.44}
\end{eqnarray}%
The calculation of the convergence constant in Eq. (\ref{1.44}) is seen to
be reduced to a one-dimensional integration over $u$.

\subsection{Higher-order cumulants and the linked-cluster theorem}

In principle, to calculate or estimate the asymptotic behavior of the higher-order  cumulants $\mu_{ck}$, $k\ge4$, one can use Eq. (\ref{1.7.1}) as used in the previous sections to examine $\mu_{c2}$ and $\mu_{c3}$. However, with increasing cumulant order the number of terms to be estimated in Eq. (\ref{1.7.1}) grows rapidly. Moreover, as we found in previous sections there are certain rules associated with connected diagrams, which help identify the terms that give non-zero contributions to the cumulants. The purpose of the present section is to generalize the results obtained for the second and third-order cumulants to higher orders.  In particular, we derive asymptotic estimates for the higher order cumulants.

By inspection of the results obtained for the first three moments $\mu
_{p},$ $p=1,2,3$, Eqs. (\ref{1.10.2}), (\ref{1.13}), (\ref{1.14}), (\ref%
{1.23}), and (\ref{1.24}), we can generalize the integral expressions for the moments to an arbitrary index $p$:%
\begin{equation}
\mu _{p}=\int\limits_{0}^{1}\cdots \int\limits_{0}^{1}du_{1}\cdots
du_{p}g_{p}(u_{1},\hat{d}_{1},\cdots ,u_{p},\hat{d}_{p})V(1)\cdots V(p)
\label{1.45}
\end{equation}%
where%
\begin{eqnarray}
g_{p}(u_{1},\hat{d}_{1},\cdots ,u_{p},\hat{d}_{p}) &\equiv
&\int\limits_{-\infty }^{\infty }\cdots \int\limits_{-\infty }^{\infty }%
\frac{d\xi _{1}\cdots d\xi _{p}}{(2\pi )^{p/2}}\exp \left( -\frac{1}{2}%
\sum_{i=1}^{p}\xi _{i}^{2}+\sum_{k=1}^{p}\varepsilon _{k}^{1/2}\hat{d}%
_{k}\sum_{i=1}^{k}\alpha _{ki}\xi _{i}\right)  \notag \\
&=&\exp \left( \frac{1}{2}\sum_{i,j=1}^{p}\gamma _{ij}\hat{d}_{i}\hat{d}%
_{j}\right).  \label{1.46}
\end{eqnarray}%
Equation (\ref{1.46}) is derived in Appendix B.

It is useful to examine the diagonal terms in the exponent of Eq. (\ref{1.46})
\begin{eqnarray}
\prod_{i=1}^{p}\left[ g_{1}(u_{i},\hat{d}_{i})V(i)\right] =
\prod_{i=1}^{p}V_{PA}(i)  \label{1.47}
\end{eqnarray}
so that we can rewrite Eq. (\ref{1.45}) as 
\begin{eqnarray}
\mu_{p}=\int\limits_{0}^{1}\cdots\int\limits_{0}^{1} du_{1}\cdots
du_{p}\,f_{p}\left(\left\lbrace \Gamma_{ij}\right\rbrace _{i<j=1}^{p}\right)
V_{PA}(1)\cdots V_{PA}(p)  \label{1.48}
\end{eqnarray}
where 
\begin{eqnarray}
f_{p}\left(\left\lbrace \Gamma_{ij}\right\rbrace _{i<j=1}^{p} \right)=
\exp\left( \sum_{i<j=1}^{p}\Gamma_{ij}\right).  \label{1.49}
\end{eqnarray}
The total number of couplings $\Gamma_k\equiv\Gamma_{ij}$, where $i<j$ and $%
i,j=1,\ldots,p$, $p\ge 2$, between all possible pairs of $p$ vertices is 
\begin{eqnarray}
K_p\equiv C_{p}^{2}=\dfrac{p(p-1)}{2}.  \label{1.50}
\end{eqnarray}

We now define
$L_{1}$ and $L_{2}$ to be an arbitrary partition among $p$ vertices such
that the first cluster contains $k_{1}$ vertices and the second one $k_{2}$
vertices, with the total number of vertices being $k_{1}+k_{2}=p$. Without
loss of generality, we assume that the first cluster contains vertices with
numbers from 1 to $k_{1}$, $L_1\equiv\{1,\ldots,k_1\}$, and the second cluster
contains vertices from $k_{1}+1$ to $p$, $L_2\equiv\{k_1+1,\ldots,p\}$.
Using this partition we call the exponent of Eq. (\ref{1.49}) the
``Hamiltonian'' $H_{12}$ of the system of $p$ vertices (using an analogy with a real Hamiltonian system), which can be divided as 
\begin{eqnarray}
H_{12}&=& \sum_{i<j=1}^{p}\Gamma_{ij}  \notag \\
&=&H_{1}+H_{2}+V_{12}  \label{1.51}
\end{eqnarray}
where 
\begin{eqnarray}
H_{1}&=& \sum_{i<j=1}^{k_1}\Gamma_{ij}  \notag  \\
H_{2}&=&\sum_{i<j=k_1+1}^{p}\Gamma_{ij}  \label{1.52}
\end{eqnarray}
are, respectively, the ``Hamiltonians'' of the first and second clusters.  The expression
\begin{eqnarray}
V_{12}= \sum_{i\in L_1,\,j\in L_2} \Gamma_{ij}  \label{1.53}
\end{eqnarray}
is the ``interaction'' between the clusters.

In accord with other diagrammatic approaches in physics, we define a diagram that has all vertices connected to be {\em linked}. If $L_1$
and $L_2$ are two linked diagrams, we call $L_1+L_2$ {\em linked} if the resulting diagram is linked.
In contrast, if the
interaction between the two diagrams $V_{12}= 0$ in $L_1+L_2$, the resulting diagram must be
disconnected or unlinked. In the unlinked case, the factorization
property given in Eq. (\ref{1.54}) is valid
\begin{eqnarray}
\mu_p=\mu_{k_1}\mu_{k_2}.  \label{1.54}
\end{eqnarray}
Additionally, if Eq.(\ref{1.54}) is satisfied, then diagrams corresponding to the product of two moments must be unlinked.

In previous sections we have shown that the contributions to the second and third-order cumulants are zero when the corresponding diagrams are disconnected. The cancellation of disconnected
diagrams found in the second and third cumulants is an example of the linked
cluster theorem, which applies not only in the present context, but in
quantum field theory, many-body theory and cumulant expansions in
statistical physics.\cite{kubo_generalized_1962,englert_linked_1963,arai_cumulant_1967,fox_critique_1976, becker_unknown_1990,sanyal_systematic_1993, fauser_cumulants_1996, mandal_non-perturbative_2002, negele_quantum_1987, goldenfeld_lecturesphase_1982}

\subsubsection{The linked-cluster theorem}

We now prove that only the terms that can be represented by linked diagrams make non-zero contributions to the corresponding cumulants. This statement is often called the linked-cluster theorem.  We have already checked directly this statement for the second and third-order cumulants.  Using the diagrammatic approach discussed in the previous subsection, we are now in a position to prove the linked cluster theorem for an arbitrary order cumulant term
using the method of mathematical induction. We assume
that cumulants from the second up to the $p^{th}$ order take zero values on
the corresponding disconnected diagrams. Further, without loss of generality
we assume that a disconnected $p+1$ cluster of vertices is made of two
linked clusters $L_1$ and $L_2$ of sizes $k_1$ and $k_2$, respectively, such
that $k_1+k_2=p+1$.  Consequently, the interaction between the two clusters $%
V_{12}= 0$. It is straightforward to generalize the proof to the case
of a disconnected cluster that is made of several linked clusters. In a
second induction step, we must prove that the same property remains valid
for this disconnected, $p+1$-vertex diagram; i.e. we must prove that $%
\mu_{c(p+1)}=0$. Rewriting Eq. (\ref{1.18.6}) for the 
cumulant term of order $(p+1)$, we obtain 
\begin{eqnarray}
\mu _{c(p+1)}=\mu
_{p+1}-\sum\limits_{r=2}^{p+1}\sum\limits_{S_{1},S_{2},\ldots ,S_{r}}\mu
_{c} \left[ S_{1}\right] \mu _{c}\left[ S_{2}\right] \ldots \mu _{c}\left[
S_{r}\right].  \label{1.55}
\end{eqnarray}
In view of the factorization property Eq. (\ref{1.54}), we have 
\begin{eqnarray}
\mu _{p+1}=\mu_{k_1}\mu_{k_2}  \label{1.56}
\end{eqnarray}
The summation in Eq. (\ref{1.55}) 
\[
\sum\limits_{r=2}^{p+1}\sum\limits_{S_{1},S_{2},\ldots,S_{r}}
\]
covers all possible ``proper'' partitions $S_{1},S_{2},\ldots,S_{r}$, $r\ge 2$
among $p+1$ vertices. Some members of these partitions, that we identify using the notation $%
S_{disc}\in \{1,\ldots,p+1\}$, contain vertices from both clusters, $%
L_1\equiv\{1,\ldots,k_1\}$ and $L_2\equiv\{k_1+1,\ldots,p+1\}$. Consequently,
diagrams corresponding to these partition members are disconnected.
Moreover, because $r\ge 2$ the maximum size of $S_{disc}$, the maximum
possible number of vertices in $S_{disc}$, is less or equal to $p$, and
 by induction, we have $\mu_{c}\left[S_{disc} \right]=0 $. We then find that non-zero contributions to the sum in Eq. (\ref{1.55}) are expected
to arise only from partition clusters that belong either to $L_{1}$ or $L_{2}$. Equation (\ref{1.55}) can be rewritten as 
\begin{eqnarray}
\mu _{c(p+1)}=\mu_{k_1}\mu_{k_2}&-&
\sum\limits_{l_{1}=1}^{k_{1}}\sum\limits_{S_{1}^{(1)},%
\ldots,S_{l_{1}}^{(1)}} \mu _{c}\left[ S_{1}^{(1)}\right] \ldots \mu _{c}%
\left[ S_{l_{1}}^{(1)}\right]  \notag \\
&\times&
\sum\limits_{l_{2}=1}^{k_{2}}\sum\limits_{S_{1}^{(2)},%
\ldots,S_{l_{2}}^{(2)}} \mu _{c}\left[ S_{1}^{(2)}\right] \ldots \mu _{c}%
\left[ S_{l_{2}}^{(2)}\right]  \label{1.57}
\end{eqnarray}
where $S_{1}^{(1)},\ldots,S_{l_{1}}^{(1)}$ and $S_{1}^{(2)},%
\ldots,S_{l_{2}}^{(2)}$ are, repectively, partitions of $L_1$ and $L_2$.
Finally, from the representation given in Eq.(\ref{1.18.5}), it follows that $\mu
_{c(p+1)}=0$. The linked-cluster theorem is then proved.

\subsubsection{Asymptotic convergence rates for the truncated cumulant expansion}

According to the linked-cluster theorem, the expression given in Eq. (\ref{1.7.1}) for cumulants
includes only the terms that correspond to linked or connected
diagrams. For this reason, we must omit terms in (\ref{1.7.1}) having $%
r\ge 2$, the terms made of products of moments, whose diagrams are
disconnected. The $p^{th}$ order cumulant can then be written as 
\begin{eqnarray}
\mu_{cp}=\int\limits_{0}^{1}\cdots\int\limits_{0}^{1} du_{1}\cdots
du_{p}\,f_{cp}\left(\left\lbrace \Gamma_{k}\right\rbrace _{k=1}^{K_p}
\right)V_{PA}(1)\cdots V_{PA}(p)  \label{1.58}
\end{eqnarray}
where 
\begin{eqnarray}
f_{cp}&=&\exp\left(\sum_{k=1}^{K_p}\Gamma_{k}\right)_c  \notag \\
&=&\sum_{k=0}^{\infty}\dfrac{\left(\sum\limits_{k=1}^{K_p}\Gamma_k\right)_c^k%
}{k!}  \notag \\
&=&\sum_{k=p-1}^{\infty}\sum_{p_1+\cdots+p_{K_p}=k} \dfrac{%
\left(\Gamma_1^{p_1}\cdots\Gamma_{K_p}^{p_{K_p}}\right)_c} {p_1!\cdots
p_{K_p}!}  \label{1.59}
\end{eqnarray}
Here, the subscrit $c$ equivalently denotes ``cumulant'' and
``connected''. The inner summation over non-negative integers $%
p_1,p_2,\ldots,p_{K_p}$ is performed under the restrictions that $%
p_1+p_2+\ldots+p_{K_p}=k$ and that all the vertices in the expansion should
be connected by $\Gamma_k\equiv \Gamma_{ij}$ couplings. The connectness of
vertices is stressed by the subscript $c$. Using the last restriction,
the outer summation over $k$, starts at $p-1$,
because $p$ vertices cannot be connected by less than $p-1$ pairwise lines
or couplings.

The lowest order contribution to Eq. (\ref{1.59}) is the term $k=p-1$
\begin{equation}
f_{cp}=\sum_{\mathrm{linked\;diagrams}}\Gamma _{k_{1}}\Gamma _{k_{2}}\cdots
\Gamma _{k_{p-1}}  \label{1.59.1}
\end{equation}
where $\Gamma _{k_{1}},\Gamma _{k_{2}},\ldots ,\Gamma _{k_{p-1}}$ are
pairwise couplings such that all $p$ vertices are linked. Among all possible vertex connections
we distinguish consecutive and centered
ones. Figure \ref{fig:two} (c) illustrates the notion of consecutive, centered, and loop diagrams in case of $p=4$.
Consecutive connections are those whose vertices, ordered in an
arbitrary way, are linked consecutively one after another. If we label a set of
ordered vertices from 1 to $p$, a consecutive connection
corresponds to the following chain of operators 
\begin{equation}
\Gamma _{12}\Gamma _{23}\cdots \Gamma _{(p-1)p}.  \label{1.59.2}
\end{equation}
If a vertex is
linked to all other $p-1$ vertices, then such a connection will be called
{\em centered}. The corresponding ``centered to vertex 1'' operator chain
takes the form 
\begin{equation}
\Gamma _{12}\Gamma _{13}\cdots \Gamma _{1p}  \label{1.59.3}
\end{equation}
For a consecutively linked cluster [Eq. (\ref{1.59.2})], the contribution to
the cumulant term is
\begin{eqnarray}
\mu _{cp}^{cons} &=&\sigma ^{2(p-1)}\int\limits_{0}^{1}\cdots
\int\limits_{0}^{1}du_{1}\cdots du_{p}\,\gamma _{12}\gamma _{23}\cdots
\gamma _{(p-1)p}  \notag \\
&\times &\dfrac{dV_{PA}(1)}{dx}\dfrac{d^{2}V_{PA}(2)}{dx^{2}}\dfrac{%
d^{2}V_{PA}(3)}{dx^{2}}\cdots \dfrac{d^{2}V_{PA}(p-1)}{dx^{2}}\dfrac{%
dV_{PA}(p)}{dx}  \label{1.63}
\end{eqnarray}%
The numerical evaluation of the $p$-dimensional integral Eq.(\ref%
{1.63}) with the Gauss-Legendre quadrature formula requires in total $%
2N_{q}$ calls to calculate the first and second derivatives of the potential
function $V_{PA}$, where $N_{q}$ is the number of quadrature points in a
single dimension.

In Appendix A, we derive an asymptotic estimate [Eq. (\ref{A34})] for integrals
of the type Eq. (\ref{A24}), from which we obtain an estimate for%
\begin{equation}
\mu _{cp}^{cons}\sim\frac{C_{p}^{cons}}{N^{2p-1}}  \label{1.64}
\end{equation}%
where $C_{p}^{cons}$ is a convergence constant. Moreover, in Appendix A we
estimate the contribution[Eq. (\ref{A35})] from the loop diagram%
\begin{equation}
\Gamma _{12}\Gamma _{23}\cdots \Gamma _{(p-1)p}\Gamma _{p1},  \label{1.65}
\end{equation}%
which includes $p$ $\Gamma $-couplings. We find that the corresponding loop
diagram contributes
\begin{eqnarray}
\mu _{cp}^{loop} &=&\sigma ^{2p}\int\limits_{0}^{1}\cdots
\int\limits_{0}^{1}du_{1}\cdots du_{p}\,\gamma _{12}\gamma _{23}\cdots
\gamma _{(p-1)p}\gamma _{p1}  \notag \\
&&\times \left[ \prod\limits_{i=1}^{p}\dfrac{d^{2}V_{PA}(i)}{dx^{2}}\right]
\label{1.66}
\end{eqnarray}
to the cumulant.  According to Eq.(\ref{A38}), Eq. (\ref{1.66})
is of the same order as the consecutive term; i.e.%
\begin{equation}
\mu _{cp}^{loop}\sim\frac{C_{p}^{loop}}{N^{2p-1}}  \label{1.67}
\end{equation}

For a centered diagram representing Eq. (\ref{1.59.3}) we obtain%
\begin{eqnarray}
\mu _{cp}^{cent} &=&\sigma ^{2(p-1)}\int\limits_{0}^{1}\cdots
\int\limits_{0}^{1}du_{1}\cdots du_{p}\,\gamma _{12}\gamma _{13}\cdots
\gamma _{1p}  \notag \\
&&\times \dfrac{d^{p-1}V_{PA}(1)}{dx^{p-1}}\left[ \prod\limits_{i=2}^{p}%
\dfrac{dV_{PA}(i)}{dx}\right]   \label{1.68}
\end{eqnarray}%
In Appendix A, we estimate this type of integral [(\ref{A42})] finding
\begin{equation}
\mu _{cp}^{cent}\sim\frac{C_{p}^{cent}}{N^{2p-1}}  \label{1.69}
\end{equation}%
Collecting all estimates Eqs. (\ref{1.64}), (\ref{1.67}%
), and (\ref{1.69}) we can conclude that the total $p^{th}$ cumulant term
scales as%
\begin{equation}
\mu _{cp}\sim\frac{C_{p}}{N^{2p-1}},  \label{1.70}
\end{equation}
From Eq.(\ref{1.70}), one can obtain an asymptotic estimate for the convergence rate of the cumulant expansion. If the cumulant expansion is truncated at order $p$, then the convergence rate of the truncated cumulant expansion is defined by the asymptotic behavior of the next cumulant term of order $(p+1)$.   The resulting term scales as  $N^{-(2p+1)}$, a central result of this work. 

\section{Discussion}

We have developed an asymptotic analysis of the cumulant
expansion for the Fourier path integral representation of the quantum, imaginary-time
density matrix. Starting from the Feynman-Kac formula expressed in Eq. (\ref{1.1}), we have used the perturbation series given in Eqs. (\ref{1.3}) and (\ref{1.4}), expanded in powers
of the path averaged potential energy defined in Eq. (\ref{1.5}). For the moments $\mu _{p}$, $p=1,2,\ldots $ generated by this
perturbation expansion we have found the compact integral representation expressed in Eq. (\ref
{1.48}). The integrand of the $p$-dimensional integral given in Eq. (\ref{1.48}) is
defined as the exponential function [Eq.(\ref{1.49})] of a linear combination of
differential operators $\Gamma _{ij}$ acting on the product of $p$ potential
functions $V_{PA}(1),\cdots V_{PA}(p)$. The operator $\Gamma _{ij}$ is proportional to
the $\gamma _{ij}$ function [Eq.(\ref{GS6})], which asymptotically behaves as $%
\gamma _{i \ne j}\sim 1/N^3$, which in the limit $N\rightarrow \infty $ goes to 0.  In the same limit the exponential
operator function given in Eq. (\ref{1.49}) reduces to the identity operator.  Using the relation $\varepsilon \sim 1/N$, we can conclude that $V_{PA}$, defined
in Eq. (\ref{1.10.3}), tends to $V$.
Consequently, in the limit of large $N$ we obtain a factorization of $\mu _{p}\rightarrow %
\left[ \bar{V}\right] ^{p}$, which results in Taylor's series expansion for
the original exponential representation given in Eq.~(\ref{1.4}).

When calculating the Feynman-Kac formula, it would be impractical
to use the perturbation series truncated at $p=p_{max}$,
because $\mu _{p}$ considered as a function of $N$ does not go to zero as $%
N\rightarrow \infty $.  As we have demonstrated, the cumulant expansion given in Eq.~(\ref%
{1.7}) behaves well asymptotically.  In particular the $p^{th}$-order cumulant term $\mu _{cp}$ scales as $N^{-(2p-1)}$. At large enough 
$N$,  the truncated cumulant expansion is
expected to provide a good approximation to the exact density matrix, with the error
scaling as $N^{-(2p_{\max }+1)}$.

Using the polynomial expansion in products $\Gamma _{1}^{p_{1}}\cdots \Gamma
_{K_{p}}^{p_{K_{p}}}$\ of powers of $\Gamma _{k}=\Gamma _{ij}=\gamma _{ij}%
\hat{d}_{i}\hat{d}_{j}$ for the exponential operator function given in Eq. (\ref{1.59}),
the integrand of the $p$-dimensional integral given in Eq. (\ref{1.58}) can be
represented as a sum of products of $p$ potential functions $%
V_{PA}(1),\cdots V_{PA}(p)$ and their derivatives times the corresponding polynomial of $\gamma$-functions. Only the linked diagrams corresponding to the chains of operators $\Gamma
_{1}^{p_{1}}\cdots \Gamma _{K_{p}}^{p_{K_{p}}}$ contribute to the final result.  A numerical evaluation of the $p$-dimensional integral can be performed using,
e.g. a $p$-dimensional Gauss-Legendre quadrature formula, with $N_{q}$
quadrature points.

To find an optimal numerical scheme to evaluate the cumulants, it is important to estimate the amount of numerical work required. A reasonable estimate of this time 
is the number of potential function calls $N_{call}$ required to
compute the integral given in Eq. (\ref{1.58}). To minimize
the time required to calculate the potential energy function at a fixed
argument, it is useful to have an analytic expression for the $V_{PA}$
function. It is well known that the Gaussian transform expressed in Eq. (\ref{eq:paap}) can be evaluated analytically either for a polynomial or a Gaussian-type potential
function. In general, we can assume that $V$ can be fitted by a finite
combination of polynomial and/or Gaussian-type potential functions. For
example, we have shown that a fit to the Lennard-Jones potential using two Gaussians gives numerical results that are indistinguishable to within statistical fluctuations of Monte Carlo path integral simulations\cite{kunikeev_numerical_2009}. 

As suggested in the previous paragraph, the computational cost in path integral simulations is dominated by the computational overhead required to evaluate the potential energy.  Efficiency gains resulting from improved asymptotic convergence rates must be balanced with any possible increase in the number of calls required to evaluate the system potential.  We have shown in Eq. (\ref{1.58}) that we can attain the asymptotic convergence behavior at a given cumulant order $p$ without including all terms in the series expansion for the cumulant.  By truncating the expansion we are able to minimize the number of potential energy evaluations without sacrificing the improved asymptotic convergence rate.  An important example is the case of the second-order cumulant discussed in the main text.  We can attain the $N^{-5}$ asymptotic convergence rate by including only the two terms represented in Eqs. (\ref{1.19.1}) and (\ref{1.19.5}).  By virtue of the truncation,
$N_{call}=3N_{q}$, resulting in a scaling that is linear in $N_q$.  As a result of Eq. (\ref{1.58}) this linear scaling in $N_q$ is maintained at all cumulant orders $p$ so that $N_{call}=(r_{\max }+1)N_{q}$ where $r_{\max }$ is the maximum order of the derivative of the potential
function $V_{PA}$ needed in the expansion to ensure the asymptotic convergence rate is $N^{-(2p+1)}$.  Even though the current paper has examined the simple case of one-dimensional systems, by using the expansion, the remarkable linear scaling in $N_q$ is maintained even for many-particle systems.  Of course, if $N_q$ must be increased with cumulant order to obtain sufficiently accurate evaluations of the integrals with respect to the imaginary time variable, the scaling with cumulant order might be more severe than linear.  Only numerical experience will enable a full understanding of the scaling with $N_q$.  However, from the purely formal results, we expect to be able to extend partial averaging to higher order cumulants in a variety of important quantum problems.

In reference \onlinecite{kunikeev_numerical_2009}, we have numerically investigated convergence
characteristics for the energy and heat capacity calculated with the first
cumulant term in a one-dimensional Lennard-Jones model. Effects of the
higher order cumulant terms and their convergence properties for $N$-body systems are the subject of a
separate publication.

\section*{Acknowledgements}

This work has been supported in part by National Science Foundation Grant Number CHE0554922.  JDD gratefully acknowledges grant support of this research through the DOE Multiscale Mathematics and Optimization for Complex Systems program.

\appendix
\section{Asymptotic estimates for integrals}

The purpose of this Appendix is to derive the asymptotic behavior of the integrals encounted in the main text.
First, we examine the behavior of the two-time integral defined in Eq.(\ref{1.6.5}) in the asymptotic limit ($N\rightarrow \infty $)
\begin{eqnarray}
I_{2}(k) &=&\int\limits_{0}^{1}\int\limits_{0}^{1}du_{1}du_{2}\left[ \gamma
(u_{1},u_{2})\right] ^{k}f_{1}(u_{1})f_{2}(u_{2})  \notag \\
&=&2^{k}\sum_{n_{1},\cdots ,n_{k}=N+1}^{\infty }\dfrac{1}{(\pi
n_{1})^{2}\cdots (\pi n_{k})^{2}}\int_{0}^{1}du_{1}\sin (\pi
n_{1}u_{1})\cdots \sin (\pi n_{k}u_{1})f_{1}(u_{1})  \notag \\
&\times &\int_{0}^{1}du_{2}\sin (\pi n_{1}u_{2})\cdots \sin (\pi
n_{k}u_{2})f_{2}(u_{2})  \label{A1}
\end{eqnarray}%
where $f_{1}(u)$ and $f_2(u)$ are smooth functions of $u$ and $k\geq 1$ is an integer. In
the second line we have used expansion given in Eq. (\ref{1.18}) for $\gamma $. 
Expanding both $f_1$ and $f_2$ (denoted generally by $f_{1,2}$) in a Fourier series we obtain 
\begin{equation}
f_{1,2}(u_{1,2})=\sum_{m_{1,2}=0}^{\infty }\widetilde{f}_{1,2}(m_{1,2})\cos
(\pi m_{1,2}u_{1,2}).  \label{A2}
\end{equation}
If Eq.(\ref{A2}) is substituted into Eq.(\ref{A1}), the trigonometric integrals can be evaluated analytically resulting in the expression
\begin{eqnarray}
I_{2}(k) &=&2^{k}\sum_{n_{1}\cdots n_{k}=N+1}^{\infty }\dfrac{1}{(\pi
n_{1})^{2}\cdots (\pi n_{k})^{2}}\sum_{m_{1},m_{2}=0}^{\infty }\widetilde{f}%
_{1}(m_{1})\widetilde{f}_{2}(m_{2})  \notag \\
&\times &J(m_{1},n_{1},\cdots ,n_{k})J(m_{2},n_{1},\cdots ,n_{k})  \label{A3}
\end{eqnarray}%
where 
\begin{eqnarray}
J(n_{0},n_{1},\cdots ,n_{k}) &\equiv &\int\limits_{0}^{1}du\cos (\pi
n_{0}u)\sin (\pi n_{1}u)\cdots \sin (\pi n_{k}u)  \notag \\
&=&\frac{1}{2^{k+1}i^{k}}\sum_{\sigma _{0},\sigma _{1},\cdots ,\sigma
_{k}=\pm 1}\sigma _{1}\cdots \sigma _{k}\left\{ 
\begin{array}{c}
1\quad \mathrm{if}\quad \nu =0 \\ 
\dfrac{(-1)^{\nu }-1}{i\pi \nu }\quad \mathrm{if}\quad \nu \neq 0%
\end{array}%
\right. ,  \label{A4} \\
\nu &=&\sum\limits_{j=0}^{k}n_{j}\sigma _{j}  \notag
\end{eqnarray}
We assume that $f_{i}$, $i=1,2$ are quadratically integrable. For such functions, to order
$\varepsilon$, these functions are well represented by a
finite series with $M_{\varepsilon 1,2}$  terms
\begin{equation}
\left[ \int\limits_{0}^{1}du\left(
f_{1,2}(u)-\sum\limits_{m=0}^{M_{\varepsilon 1,2}}\widetilde{f}_{1,2}(m)\cos
(\pi mu)\right) ^{2}\right] ^{1/2}\leqslant \varepsilon .  \label{A4.1}
\end{equation}
In the asymptotic limit we recognize the condition $N>M_{\varepsilon 1,2}$%
. Because the integral in Eq. (\ref{A4}) is real, we observe the separate contributions for even and odd values of $k$.
At $k=1$ and $2$ we obtain, respectively,%
\begin{equation}
J(n_{0},n_{1})=\frac{1-(-1)^{n_{0}+n_{1}}}{(\pi n_{1})\left(
1-(n_{0}/n_{1})^{2}\right) }=\frac{1-(-1)^{n_{0}+n_{1}}}{\pi n_{1}}%
\sum_{j=0}^{\infty }\left( \frac{n_{0}}{n_{1}}\right) ^{2j}  \label{A5}
\end{equation}%
and%
\begin{equation}
J(n_{0},n_{1},n_{2})=\frac{1}{8}\left\{ \delta _{n_{1},n_{2}+n_{0}}+\delta
_{n_{1},n_{2}-n_{0}}+\delta _{n_{2},n_{1}+n_{0}}+\delta
_{n_{2},n_{1}-n_{0}}\right\}  \label{A6}
\end{equation}%
where $\delta _{n,m}$ is the Kronecker delta. After substituting Eq. (\ref{A5}%
) into (\ref{A3}), we obtain 
\begin{eqnarray}
I_{2}(1) &=&\sum_{n=N+1}^{\infty }\dfrac{2}{(\pi n)^{4}}%
\sum_{j_{1},j_{2}=0}^{\infty }\dfrac{(-1)^{j_{1}+j_{2}}}{(\pi
n)^{2(j_{1}+j_{2})}}\left[ (-1)^{n}f_{1}^{(2j_{1})}(1)-f_{1}^{(2j_{1})}(0)%
\right]  \notag \\
&\times &\left[ (-1)^{n}f_{2}^{(2j_{2})}(1)-f_{2}^{(2j_{2})}(0)\right] ,
\label{A7}
\end{eqnarray}%
where the following equalities have been used 
\begin{eqnarray}
\sum_{m=0}^{M_{\varepsilon }}(-1)^{m}\widetilde{f}(m)m^{2j} &=&\dfrac{%
(-1)^{j}}{\pi ^{2j}}f^{(2j)}(1),  \label{A7.1} \\
\sum_{m=0}^{M_{\varepsilon }}\widetilde{f}(m)m^{2j} &=&\dfrac{(-1)^{j}}{\pi
^{2j}}f^{(2j)}(0).  \label{A7.2}
\end{eqnarray}%
The functions on the right-hand side of Eqs.  (\ref{A7.1}) and  (\ref{A7.2}) are $\varepsilon $-approximants using a truncated Fourier series.
The leading term in Eq. (\ref{A7})
is seen to be of third order 
\begin{eqnarray}
I_{2}(1) &=&\sum_{n=N+1}^{\infty }\dfrac{2}{(\pi n)^{4}}\left[
(-1)^{n}f_{1}(1)-f_{1}(0)\right]  \notag \\
&\times &\left[ (-1)^{n}f_{2}(1)-f_{2}(0)\right] +O(N^{-5}).  \label{A8}
\end{eqnarray}%
Equation (\ref{A8}) can be rewritten 
\begin{eqnarray}
I_{2}(1) &=&\dfrac{2}{\pi ^{4}}\left\{ \dfrac{1}{3N^{3}}\left[
f_{1}(0)f_{2}(0)+f_{1}(1)f_{2}(1)\right] \right.  \notag \\
&-&\left. S_{4}(N)\left[ f_{1}(1)f_{2}(0)+f_{1}(0)f_{2}(1)\right] \right\} +O(N^{-5}) ,
\label{A8.1}
\end{eqnarray}%
where 
\begin{equation}
S_{k}(N)\equiv \sum_{n=N+1}^{\infty }\dfrac{(-1)^{n}}{n^{k}}.  \label{A8.2}
\end{equation}%
The asymptotic behavior in $N$ can be estimated separately for even and odd values of $N$. At even $N$, we have 
\begin{eqnarray}
S_{k}(N) &=&\sum_{n=1}^{\infty }\dfrac{1}{(N+2n)^{k}}-\sum_{n=0}^{\infty }%
\dfrac{1}{(N+2n+1)^{k}}  \notag \\
&=&\dfrac{1}{N^{k-1}}\left\{ \sum_{n=1}^{\infty }\dfrac{1/N}{\left( 1+2%
\dfrac{n}{N}\right) ^{k}}-\sum_{n=0}^{\infty }\dfrac{1/N}{\left( 1+\dfrac{%
2n+1}{N}\right) ^{k}}\right\}  \notag \\
&\approx &\dfrac{1}{N^{k-1}}\left\{ \int_{1/N}^{\infty }\dfrac{dx}{(1+2x)^{k}%
}-\int_{0}^{\infty }\dfrac{dx}{(1+\frac{1}{N}+2x)^{k}}\right\} =-\dfrac{1}{%
2N^{k}},  \label{A8.3}
\end{eqnarray}%
whereas at $N$ odd, one finds the same estimate by modulus, but with the
opposite, \textquotedblleft $+$\textquotedblright\ sign. Consequently, at any large $%
N$ we obtain 
\begin{equation}
S_{k}(N)=\dfrac{(-1)^{N+1}}{2N^{k}}.  \label{A8.4}
\end{equation}
After substituting Eq. (\ref{A6}) into (\ref{A3}), we obtain 
\begin{eqnarray}
I_{2}(2) &=&\sum_{n=N+1}^{\infty }\dfrac{1}{(\pi n)^{4}}\left\{ \widetilde{f}%
_{1}(0)\widetilde{f}_{2}(0)+\frac{1}{2}\sum_{m=1}^{M_{\varepsilon }}%
\widetilde{f}_{1}(m)\widetilde{f}_{2}(m)\dfrac{1+\left( \frac{m}{n}\right)
^{2}}{\left[ 1-\left( \frac{m}{n}\right) ^{2}\right] ^{2}}\right\}  \notag \\
&\approx &\sum_{n=N+1}^{\infty }\dfrac{1}{(\pi n)^{4}}\left\{ \widetilde{f}%
_{1}(0)\widetilde{f}_{2}(0)+\frac{1}{2}\sum_{m=1}^{M_{\varepsilon }}%
\widetilde{f}_{1}(m)\widetilde{f}_{2}(m)\right. \notag\\
&+& 
\left.\dfrac{3}{2n^{2}}%
\sum_{m=1}^{M_{\varepsilon }}m^{2}\widetilde{f}_{1}(m)\widetilde{f}%
_{2}(m)\right\}   \label{A9}
\end{eqnarray}%
where $M_{\varepsilon }=\min (M_{\varepsilon 1},M_{\varepsilon 2})$.
After integrating by parts, we find
\begin{eqnarray}
K_{j} &\equiv
&\int_{0}^{1}du\,f_{1}^{(2j)}(u)f_{2}(u)=\int_{0}^{1}du%
\,f_{1}(u)f_{2}^{(2j)}(u)  \notag \\
&=&\delta _{j,0}\widetilde{f}_{1}(0)\widetilde{f}_{2}(0)+\frac{1}{2}%
\sum_{m=1}^{M_{\varepsilon }}(\pi m)^{2j}\widetilde{f}_{1}(m)\widetilde{f}%
_{2}(m),  \label{A10}
\end{eqnarray}%
where $j=0,1,\cdots $. Using Eq.(\ref{A10}), Eq. (\ref{A9}) can be rewritten as
\begin{equation}
I_{2}(2) =K_{0}\sum_{n=N+1}^{\infty }\dfrac{1}{(\pi n)^{4}}%
+3K_{1}\sum_{n=N+1}^{\infty }\dfrac{1}{(\pi n)^{6}}+\cdots ,  \label{A11}
\end{equation}
with
\begin{equation}
K_{0} =\int\limits_{0}^{1}du\,f_{1}(u)f_{2}(u),\quad
\end{equation}
and
\begin{equation}
K_{1}=\int\limits_{0}^{1}du\,f_{1}^{(2)}(u)f_{2}(u).
\end{equation}
We observe that 
\begin{equation}
\int\limits_{0}^{1}\int\limits_{0}^{1}du_{1}du_{2}\gamma
^{2}(u_{1},u_{2})=\sum\limits_{n=N+1}^{\infty }\frac{1}{(\pi n)^{4}}\sim 
\frac{1}{3\pi ^{4}N^{3}},
\end{equation}%
so that the leading term of (\ref{A11}) is of the third order, while
the next term is of the fifth order. Moreover, as $N\rightarrow \infty $ the
\textquotedblleft normalized\textquotedblright\  function $\gamma ^{2}(u_{1},u_{2})$
function becomes a Dirac delta function 
\begin{equation}
\lim_{N\rightarrow \infty }\frac{\gamma ^{2}(u_{1},u_{2})}{%
\int\limits_{0}^{1}\int\limits_{0}^{1}du_{1}du_{2}\gamma ^{2}(u_{1},u_{2})}%
=\delta (u_{1}-u_{2}).  \label{A13}
\end{equation}
Equation (\ref{A13}) has been proved
in Appendix B of Reference \onlinecite{predescu_asymptotic_2003}.

For $k>2$ the number of terms in Eq. (\ref{A4}) grows
exponentially with $k$, and the corresponding expressions for asymptotic constants quickly become quite unwieldy.
However, we can still obtain asymptotic estimates at $%
k\geqslant 3$ by replacing the summations involved with integrations. By virtue of the Euler-MacLaurin summation formula,\cite{abramowitz_handbook_1965} these replacements become exact in the limit that $N \rightarrow \infty$.  To demonstrate the approach, we examine Eq. (\ref{A3}) noticing that the summation indices
$n_{1},\cdots ,n_{k}$ begin with $N+1$. We next replace these summation indices 
with continuous variables $x_{1}=n_{1}/(N+1),\cdots ,x_{k}=n_{k}/(N+1)$ and approximately replace the
summation over indices by an integration over the corresponding $x$-variables
with the integration range being defined from $1$ to $\infty $.  For odd values of $k=2p+1$, $p=0,1,\cdots $, the $J$-functions in Eq. (\ref%
{A3}) are inversely proportional to $\Delta \sim N$, so that asymptotically
we obtain%
\begin{equation}
I_{2}(k=2p+1)\sim \frac{1}{N^{2p+3}}\int\limits_{1}^{\infty }\cdots
\int\limits_{1}^{\infty }\frac{dx_{1}\cdots dx_{k}}{x_{1}^{2}\cdots x_{k}^{2}%
}\varphi _{k}(x_{1},\cdots ,x_{k})=\frac{C_{k}}{N^{k+2}}.  \label{A14}
\end{equation}%
Here, there is no need to specify the function $\varphi _{k}$; the only
requirement is the finiteness of the integral.
At even values of $k=2p+2$, $p=0,1,\cdots $, the $J$-functions in Eq. (\ref{A3})
are proportional to Kronecker deltas, which asymptotically tend
to Dirac delta functions as $N\rightarrow \infty $. As a result, we obtain $k-1$
independent integration variables, which we set to be $x_{2},\cdots ,x_{k}$, with $x_{1}$
being a function of these $k-1$ variables. The asymptotic expression for the
integral takes the form
\begin{equation}
I_{2}(k=2p+2)\sim \frac{1}{N^{2p+3}}\int\limits_{1}^{\infty }\cdots
\int\limits_{1}^{\infty }\frac{dx_{2}\cdots dx_{k}}{x_{1}^{2}\cdots x_{k}^{2}%
}\varphi _{k}(x_{1},\cdots ,x_{k})=\frac{C_{k}}{N^{k+1}}.  \label{A15}
\end{equation}

We next examine the asymptotic behavior of the three-time integrals of the type encounted in Eq. (\ref{1.43}%
)%
\begin{equation}
I_{3}(1,1,1)=\iiint\limits_{0}^{1}du_{1}du_{2}du_{3}f(u_{1})f(u_{2})f(u_{3})%
\gamma (u_{1},u_{2})\gamma (u_{2},u_{3})\gamma (u_{1},u_{3})  \label{A16}
\end{equation}%
and%
\begin{equation}
I_{3}(0,1,1)=\iiint%
\limits_{0}^{1}du_{1}du_{2}du_{3}f_{1}(u_{1})f_{2}(u_{2})f_{2}(u_{3})\gamma
(u_{1},u_{2})\gamma (u_{1},u_{3})  \label{A17}
\end{equation}
Using the Fourier expansion of Eq. (\ref{A2}), we obtain asymptotic results accurate to
fifth order
\begin{eqnarray}
I_{3}(1,1,1) &=&\frac{1}{8}\sum_{m_{1},m_{2},m_{3}=0}\tilde{f}(m_{1})\tilde{f%
}(m_{2})\tilde{f}(m_{3})\sum_{n_{1},n_{2},n_{3}=N+1}\frac{1}{(\pi
n_{1})^{2}(\pi n_{2})^{2}(\pi n_{3})^{2}}  \notag \\
&&\times (\delta _{n_{2},n_{3}+m_{1}}+\delta _{n_{2},n_{3}-m_{1}})(\delta
_{n_{3},n_{1}+m_{2}}+\delta _{n_{3},n_{1}-m_{2}})(\delta
_{n_{2},n_{1}+m_{3}}+\delta _{n_{2},n_{1}-m_{3}})  \notag \\
&\sim &\frac{1}{5\pi ^{6}N^{5}}\left\{ \frac{3}{4}\sum_{m_{1},m_{2}=0}%
\tilde{f}(m_{1}+m_{2})\tilde{f}(m_{1})\tilde{f}(m_{2})+\frac{1}{4}\left[ 
\tilde{f}(0)\right] ^{3}\right\}  \label{A18}
\end{eqnarray}%
and%
\begin{eqnarray}
I_{3}(0,1,1) &=&\sum_{m_{1},m_{2},m_{3}=0}\tilde{f}_{1}(m_{1})\tilde{f}%
_{2}(m_{2})\tilde{f}_{2}(m_{3})\sum_{n_{1},n_{2},n_{3}=N+1}\frac{1}{(\pi
n_{1})^{2}(\pi n_{2})^{2}}  \notag \\
&&\times \frac{1}{2}\left[ \delta _{n_{1},n_{2}+m_{1}}+\delta
_{n_{1},n_{2}-m_{1}}\right] \left[ 1-(-1)^{n_{1}+m_{2}}\right] \frac{n_{1}}{%
\pi \left( n_{1}^{2}-m_{2}^{2}\right) }  \notag \\
&&\times \left[ 1-(-1)^{n_{2}+m_{3}}\right] \frac{n_{2}}{\pi \left(
n_{2}^{2}-m_{3}^{2}\right) }  \notag \\
&\sim &\sum_{n=N+1}\frac{1}{(\pi n)^{6}}\sum_{m_{1},m_{2},m_{3}=0}\tilde{f}%
_{1}(m_{1})\tilde{f}_{2}(m_{2})\tilde{f}_{2}(m_{3})  \notag \\
&&\times \left[ 1-(-1)^{n+m_{1}+m_{2}}\right] \left[ 1-(-1)^{n+m_{3}}\right].
 \label{A19}
\end{eqnarray}%
With some algebra it can be verified that
\begin{equation}
\int\limits_{0}^{1}du\left[ f(u)\right] ^{3}=\frac{3}{4}\sum_{m_{1},m_{2}=0}%
\tilde{f}(m_{1}+m_{2})\tilde{f}(m_{1})\tilde{f}(m_{2})+\frac{1}{4}\left[ 
\tilde{f}(0)\right] ^{3},  \label{A20}
\end{equation}%
\begin{eqnarray}
&&\sum_{m_{1},m_{2},m_{3}=0}\tilde{f}_{1}(m_{1})\tilde{f}_{2}(m_{2})\tilde{f}%
_{2}(m_{3})\left[ 1-(-1)^{n+m_{1}+m_{2}}\right] \left[ 1-(-1)^{n+m_{3}}%
\right]  \notag \\
&&=f_{1}(0)[f_{2}(0)]^{2}+f_{1}(1)[f_{2}(1)]^{2}-(-1)^{n}f_{2}(0)f_{2}(1)[f_{1}(0)+f_{1}(1)]
\label{A21}
\end{eqnarray}
Using Eqs. (\ref{A20}) and (\ref{A21}), Eqs. (\ref{A18}) and (\ref{A19}) can be rewritten%
\begin{equation}
I_{3}(1,1,1)\sim\frac{1}{5\pi ^{6}N^{5}}\int\limits_{0}^{1}du\left[ f(u)\right]
^{3}  \label{A22}
\end{equation}%
and%
\begin{equation}
I_{3}(0,1,1)\sim\frac{1}{5\pi ^{6}N^{5}}\left[
f_{1}(0)[f_{2}(0)]^{2}+f_{1}(1)[f_{2}(1)]^{2}\right]   \label{A23}
\end{equation}

We can also obtain an asymptotic estimate for the $p$-dimensional integral
corresponding to a consecutively linked diagram 
\begin{eqnarray}
I_{p}^{cons} &=&\int\limits_{0}^{1}\cdots \int\limits_{0}^{1}du_{1}\cdots
du_{p}\,\gamma (u_{1},u_{2})\gamma (u_{2},u_{3})\cdots \gamma
(u_{p-2},u_{p-1})\gamma (u_{p-1},u_{p})  \notag \\
&\times &f_{1}(u_{1})f_{2}(u_{2})f_{2}(u_{3})\cdots
f_{2}(u_{p-2})f_{2}(u_{p-1})f_{1}(u_{p}).  \label{A24}
\end{eqnarray}%
Substituting the expansion given in Eq. (\ref{1.18}) for the $\gamma $-function, we obtain 
\begin{eqnarray}
I_{p}^{cons} &=&2\sum\limits_{n_{1},\ldots ,n_{p-1}=N+1}^{\infty }\dfrac{%
J_{1}(n_{1})J_{1}(n_{p-1})}{(\pi n_{1})^{2}\cdots (\pi n_{p-1})^{2}}  \notag
\\
&\times &J_{2}(n_{1},n_{2})J_{2}(n_{2},n_{3})\cdots J_{2}(n_{p-2},n_{p-1}),
\label{A25}
\end{eqnarray}%
where 
\begin{eqnarray}
J_{1}(n) &=&\int\limits_{0}^{1}du\,\sin (\pi nu)f_{1}(u), \\
J_{2}(n_{1},n_{2}) &=&\int\limits_{0}^{1}du\,[\cos \pi (n_{1}-n_{2})u-\cos
\pi (n_{1}+n_{2})u]f_{2}(u).  \label{A26}
\end{eqnarray}%
Assuming the indices $n,n_1,n_2\ge N+1$, $N\to\infty$ are large, one can estimate the first integral using integration by parts. The second integral can be estimated by neglecting the contribution from the highly oscillating  $\cos\pi(n_1+n_2)u$ function. We then obtain the following estimates
\begin{eqnarray}
J_{1}(n) &\sim&\dfrac{f_{1}(0)-(-1)^{n}f_{1}(1)}{\pi n},  \label{A27} \\
J_{2}(n_{1},n_{2}) &\rightarrow &J_{2}(n_{1}-n_{2})=
\int\limits_{0}^{1}du\,\cos [\pi (n_{1}-n_{2})u]f_{2}(u)  \label{A28}
\end{eqnarray}
Using the Fourier expansion [Eq. (\ref{A2})] and the property of orthogonality
of the cosine functions, the last integral can be reduced to 
\begin{equation}
J_{2}(n_{1}-n_{2})=\frac{1}{2}[1+\delta _{|n_{1}-n_{2}|,0}]\tilde{f}%
_{2}(|n_{1}-n_{2}|)  \label{A29}
\end{equation}%
Substituting Eqs. (\ref{A27}) and (\ref{A29}) into (\ref{A25}), we obtain 
\begin{eqnarray}
I_{p}^{cons} &=&2^{3-p}\sum\limits_{n_{1},\ldots ,n_{p-1}=N+1}^{\infty }%
\dfrac{[f_{1}(0)-(-1)^{n_{1}}f_{1}(1)][f_{1}(0)-(-1)^{n_{p-1}}f_{1}(1)]}{%
(\pi n_{1})^{3}(\pi n_{2})^{2}\cdots (\pi n_{p-2})^{2}(\pi n_{p-1})^{3}} 
\notag \\
&\times &[1+\delta _{|n_{1}-n_{2}|,0}]\tilde{f}_{2}(|n_{1}-n_{2}|)\cdots
\lbrack 1+\delta _{|n_{p-2}-n_{p-3}|,0}]\tilde{f}_{2}(|n_{p-2}-n_{p-1}|),
\label{A30}
\end{eqnarray}
It is convenient to replace the set of integer variables $n_{1},\ldots
,n_{p-1}$ running from $N+1$ \ to infinity to the set of diagonal and
off-diagonal integer variables%
\begin{equation}
(n_{1},n_{2},\ldots ,n_{p-1})\rightarrow (n_{1},\Delta
n_{2}=n_{2}-n_{1},\ldots ,\Delta n_{p-1}=n_{p-1}-n_{1}),  \label{A31}
\end{equation}%
so that the denominator in Eq. (\ref{A30}) in the new variables becomes
\begin{equation}
D=(\pi n_{1})^{3}(\pi \lbrack n_{1}+\Delta n_{2}])^{2}\cdots (\pi \lbrack
n_{1}+\Delta n_{p-2}])^{2}(\pi \lbrack n_{1}+\Delta n_{p-1}])^{3}
\label{A32}
\end{equation}
When the diagonal variable $n_{1}\gg |\Delta n_{2}|,\ldots ,|\Delta n_{p-1}|,$
we can set $\Delta n_{2}=\ldots =\Delta n_{p-1}=0$ and obtain to leading
order%
\begin{equation}
D=(\pi n_{1})^{2p}  \label{A33}
\end{equation}%
The arguments of functions in the second line of Eq. (\ref{A30})
depend only on the off-diagonal variables. For quadratically integrable
functions the Fourier coefficients 
$\tilde{f}_{2}(|\Delta n|)\rightarrow 0$
as $|\Delta n|\rightarrow \infty $, and to $\varepsilon $ error, the
function can be well approximated by a finite Fourier series, Eq. (\ref{A4.1}), with $M_{\varepsilon }+1$ terms. The second line of Eq. (\ref%
{A30}) makes sizable contributions to the sum only at finite values of the
off-diagonal variables $|\Delta n_{2}|,\ldots ,|\Delta n_{p-1}|,$ restricted
from above by the number $M_{\varepsilon }$. This upper bound implies that asymptoticaly at $%
N\gg M_{\varepsilon }$ the approximation for denominator, Eq. (\ref{A33}),
is justified and Eq. (\ref{A30}) can estimated to give%
\begin{equation}
I_{p}^{cons}\sim C\sum\limits_{n_{1}=N+1}^{\infty }\frac{1}{(\pi n_{1})^{2p}%
}=O\left( \frac{1}{N^{2p-1}}\right)  \label{A34}
\end{equation}%
This asymptotic estimate is consistent with Eqs. (\ref%
{A8.1}) and (\ref{A23}), obtained in the particular cases $p=2$ and $3$.

We can also examine the integral that corresponds to a loop diagram
\begin{eqnarray}
I_{p}^{loop} &=&\int\limits_{0}^{1}\cdots \int\limits_{0}^{1}du_{1}\cdots
du_{p}\,\gamma (u_{1},u_{2})\gamma (u_{2},u_{3})\cdots \gamma
(u_{p-1},u_{p})\gamma (u_{p},u_{1})  \notag \\
&\times &f(u_{1})\cdots f(u_{p})  \label{A35}
\end{eqnarray}%
Substituting the expansion of the $\gamma $-function as in Eq. (\ref{A30}), the
loop integral can be reduced to 
\begin{eqnarray}
I_{p}^{loop} &=&\sum\limits_{n_{1},\ldots ,n_{p}=N+1}^{\infty }\dfrac{1}{%
(\pi n_{1})^{2}\cdots (\pi n_{p})^{2}}  \notag \\
&\times &J(n_{1}-n_{p})J(n_{1}-n_{2})\cdots
J(n_{p-2}-n_{p-1})J(n_{p-1}-n_{p})  \label{A36}
\end{eqnarray}%
where 
\begin{equation}
J(n)=\int\limits_{0}^{1}du\,\cos (\pi nu)f(u)=\dfrac{1}{2}\left[ 1+\delta
_{|n|,0}\right] \tilde{f}(n)  \label{A37}
\end{equation}%
Here, the $\tilde{f}(n)$'s are the Fourier coefficients in the Fourier expansion of the
function $f(u)$. Using similar methods, we obtain the same asymptotic relations for 
\begin{equation}
I_{p}^{loop}=O\left( \frac{1}{N^{2p-1}}\right)  \label{A38}
\end{equation}

Similar arguments can be made to obtain the asymptotic behavior for consecutive and loop diagrams given in Eqs. (%
\ref{A34}) and (\ref{A38}). In the case of a
consecutive diagram we have two end vertices $1$
and $p$. Integration with respect to the variables associated with the end vertices, $u_{1}$ and $u_{p}$,
gives two extra factors in the denominator $D$ of Eq. (%
\ref{A30}), $(\pi n_{1})^{2}\rightarrow (\pi n_{1})^{3}$ and $(\pi
n_{p-1})^{2}\rightarrow (\pi n_{p-1})^{3}$. These limits arise from the asymptotic
estimate given in Eq. (\ref{A27}). 
For vertices not at the ends, we obtain an extra $\gamma (u_{1},u_{p})$
function coupling vertices $1$ and $p$, which adds an extra factor $(\pi
n_{p})^{2}$ to the denominator of Eq. (\ref{A36}). Using the diagonal
approximation $n_{1}=\ldots =n_{p-1}$ or $n_{1}=\ldots =n_{p}$ for
consecutive or loop diagrams, we find that in both cases, the denominator
has the same scaling as $(\pi n_{1})^{2p}$, from which the same asymptotic expressions,
Eqs. (\ref{A34}) and (\ref{A38}), follow. We have already faced the
consecutive and loop diagrams for the particular cases $p=2$ and $3,$ when we
have observed the same asymptotic behavior for the integrals $I_{2}(1)$ and $I_{2}(2)$[
compare Eqs. (\ref{A8.1}) and (\ref{A11})], and for the integrals $%
I_{3}(0,1,1)$ and $I_{3}(1,1,1)$ [compare Eqs. (\ref{A23}) and (\ref{A22})].

In the case of a centered diagram we need to estimate the integral%
\begin{eqnarray}
I_{p}^{cent} &=&\int\limits_{0}^{1}\cdots \int\limits_{0}^{1}du_{1}\cdots
du_{p}\,\gamma (u_{1},u_{2})\gamma (u_{1},u_{3})\cdots \gamma (u_{1},u_{p}) 
\notag \\
&&\times f_{1}(u_{1})\left[ \prod\limits_{i=2}^{p}f_{2}(u_{i})\right].
\label{A39}
\end{eqnarray}%
Using the asymptotic expansion for the $\gamma$ function [Eq. (%
\ref{A27})] and the Fourier expansion for function $f_{1}$, we obtain%
\begin{eqnarray}
I_{p}^{cent} &=&2^{p-1}\sum_{n_{1},\ldots ,n_{p-1}=N+1}^{\infty }\left[
\prod\limits_{i=1}^{p-1}\frac{f_{2}(0)-(-1)^{n_{i}}f_{2}(1)}{(\pi n_{i})^{3}}%
\right]  \notag \\
&&\times \sum_{m=0}^{\infty }J(m,n_{1},\ldots ,n_{p-1})\tilde{f}_{1}(m)
\label{A40}
\end{eqnarray}%
where $J(m,n_{1},\ldots ,n_{p-1})$ is defined in Eq. (\ref{A4}). 
Equation (\ref{A40}) is similar in structure to Eq. (\ref{A3}), and we use the method of
rescaled variables $x_{1}=n_{1}/(N+1),\ldots ,x_{p-1}=n_{p-1}/(N+1)$ in analogy to the method used in deriving Eqs.  (\ref{A14}) and (\ref{A15}).
Separately examining the cases for even or odd values of $p$, we find the
same asymptotic behaviors in both cases%
\begin{equation}
I_{p}^{cent}=O\left( \frac{1}{N^{2p-1}}\right)  \label{A42}
\end{equation}
\ 

\section{ Proof of Eq. (\protect\ref{1.46})}

Taking the Gaussian integral Eq. (\ref{1.46}) over $\xi _{1},\cdots ,\xi
_{p} $, we obtain 
\begin{eqnarray}
g_{p} &=&\exp \left( \dfrac{1}{2}\sum_{i=1}^{p}\left[ \sum_{k=i}^{p}%
\varepsilon _{k}^{1/2}\alpha _{ki}\hat{d}_{k}\right] ^{2}\right)  \notag \\
&=&\exp \left( \dfrac{1}{2}\sum_{i=1}^{p}\sum_{k,k^{\prime
}=i}^{p}\varepsilon _{k}^{1/2}\varepsilon _{k^{\prime }}^{1/2}\alpha
_{ki}\alpha _{k^{\prime }i}\hat{d}_{k}\hat{d}_{k^{\prime }}\right)
\label{B1}
\end{eqnarray}%
We next write the exponent of Eq. (\ref{B1}) as 
\begin{eqnarray}
\phi _{p} &\equiv &\dfrac{1}{2}\sum_{i=1}^{p}\sum_{k,k^{\prime
}=i}^{p}\varepsilon _{k}^{1/2}\varepsilon _{k^{\prime }}^{1/2}\alpha
_{ki}\alpha _{k^{\prime }i}\hat{d}_{k}\hat{d}_{k^{\prime }}  \notag \\
&=&\dfrac{1}{2}\sum_{k,k^{\prime }=1}^{p}c_{kk^{\prime }}\hat{d}_{k}\hat{d}%
_{k^{\prime }}  \label{B2}
\end{eqnarray}%
where the coefficients are defined by
\begin{equation}
c_{kk^{\prime }}\equiv \varepsilon _{k}^{1/2}\varepsilon _{k^{\prime
}}^{1/2}\sum_{i=1}^{\mathrm{min}\,(k,k^{\prime })}\alpha _{ki}\alpha
_{k^{\prime }i}.  \label{B3}
\end{equation}%
We assume temporarily that $k^{\prime }\geq k$. After the scalar
multiplication of Eq. (\ref{GS2}) by the vector $\vec{g}_{k^{\prime }}$, we
obtain 
\begin{equation}
g_{k^{\prime }k}\equiv \vec{g}_{k^{\prime }}\cdot \vec{g}_{k}=\sum_{i=1}^{k}%
\alpha _{ki}\alpha _{k^{\prime }i}  \label{B4}
\end{equation}%
Evidently, if $k\geq k^{\prime }$ Eq. (\ref{B4}) remains valid if $%
k\leftrightarrow k^{\prime }$ so that%
\begin{equation}
g_{k^{\prime }k}=g_{kk^{\prime }}=\sum_{i=1}^{\mathrm{min}\,(k,k^{\prime
})}\alpha _{ki}\alpha _{k^{\prime }i}  \label{B5}
\end{equation}%
at any $k$ and $k^{\prime }$. Consequently, we have%
\begin{equation}
c_{kk^{\prime }}=\varepsilon _{k}^{1/2}\varepsilon _{k^{\prime
}}^{1/2}g_{kk^{\prime }}\equiv \gamma _{kk^{\prime }},  \label{B6}
\end{equation}%
which proves Eq. (\ref{1.46}).

%\section*{References}

%\bibliography{PAtech}
\newpage
\section*{References}
\bibliography{2-18-10}

\begin{thebibliography}{35}
\expandafter\ifx\csname natexlab\endcsname\relax\def\natexlab#1{#1}\fi
\expandafter\ifx\csname bibnamefont\endcsname\relax
  \def\bibnamefont#1{#1}\fi
\expandafter\ifx\csname bibfnamefont\endcsname\relax
  \def\bibfnamefont#1{#1}\fi
\expandafter\ifx\csname citenamefont\endcsname\relax
  \def\citenamefont#1{#1}\fi
\expandafter\ifx\csname url\endcsname\relax
  \def\url#1{\texttt{#1}}\fi
\expandafter\ifx\csname urlprefix\endcsname\relax\def\urlprefix{URL }\fi
\providecommand{\bibinfo}[2]{#2}
\providecommand{\eprint}[2][]{\url{#2}}

\bibitem[{\citenamefont{Feynman and Hibbs}(1965)}]{feynman_quantum_1965}
\bibinfo{author}{\bibfnamefont{R.}~\bibnamefont{Feynman}} \bibnamefont{and}
  \bibinfo{author}{\bibfnamefont{A.}~\bibnamefont{Hibbs}},
  \emph{\bibinfo{title}{Quantum mechanics and path integrals}}
  (\bibinfo{publisher}{{McGraw-Hill}}, \bibinfo{address}{New York},
  \bibinfo{year}{1965}).

\bibitem[{\citenamefont{Berne and Thirumalai}(1986)}]{berne_simulation_1986}
\bibinfo{author}{\bibfnamefont{B.~J.} \bibnamefont{Berne}} \bibnamefont{and}
  \bibinfo{author}{\bibfnamefont{D.}~\bibnamefont{Thirumalai}},
  \bibinfo{journal}{Ann. Rev. Phys. Chem.} \textbf{\bibinfo{volume}{37}},
  \bibinfo{pages}{401} (\bibinfo{year}{1986}).

\bibitem[{\citenamefont{Doll et~al.}(1990)\citenamefont{Doll, Freeman, and
  Beck}}]{doll_equilibrium_1990-1}
\bibinfo{author}{\bibfnamefont{J.~D.} \bibnamefont{Doll}},
  \bibinfo{author}{\bibfnamefont{D.~L.} \bibnamefont{Freeman}},
  \bibnamefont{and} \bibinfo{author}{\bibfnamefont{T.~L.} \bibnamefont{Beck}},
  \bibinfo{journal}{Adv. Chem. Phys.} \textbf{\bibinfo{volume}{78}},
  \bibinfo{pages}{61} (\bibinfo{year}{1990}).

\bibitem[{\citenamefont{Ceperley}(1995)}]{ceperley_path_1995-1}
\bibinfo{author}{\bibfnamefont{D.~M.} \bibnamefont{Ceperley}},
  \bibinfo{journal}{Rev. Mod. Phys.} \textbf{\bibinfo{volume}{67}},
  \bibinfo{pages}{279} (\bibinfo{year}{1995}).

\bibitem[{\citenamefont{Kleinert}(1995)}]{kleinert_path_1995}
\bibinfo{author}{\bibfnamefont{H.}~\bibnamefont{Kleinert}},
  \emph{\bibinfo{title}{Path integrals in quantum mechanics, statistics and
  polymer physics}} (\bibinfo{publisher}{World Scientific},
  \bibinfo{address}{Singapore}, \bibinfo{year}{1995}).

\bibitem[{\citenamefont{Chaichian and Demichev}(2001)}]{chaichian_path_2001}
\bibinfo{author}{\bibfnamefont{M.}~\bibnamefont{Chaichian}} \bibnamefont{and}
  \bibinfo{author}{\bibfnamefont{I.}~\bibnamefont{Demichev}},
  \emph{\bibinfo{title}{Path Integrals: Stochastic processes and quantum
  mechanics}}, vol.~\bibinfo{volume}{1} (\bibinfo{publisher}{{IOP} Publishing},
  \bibinfo{address}{Bristol}, \bibinfo{year}{2001}).

\bibitem[{\citenamefont{Schweizer et~al.}(1981)\citenamefont{Schweizer, Stratt,
  Chandler, and Wolynes}}]{schweizer_convenient_1981}
\bibinfo{author}{\bibfnamefont{K.~S.} \bibnamefont{Schweizer}},
  \bibinfo{author}{\bibfnamefont{R.~M.} \bibnamefont{Stratt}},
  \bibinfo{author}{\bibfnamefont{D.}~\bibnamefont{Chandler}}, \bibnamefont{and}
  \bibinfo{author}{\bibfnamefont{P.~G.} \bibnamefont{Wolynes}},
  \bibinfo{journal}{J. Chem. Phys.} \textbf{\bibinfo{volume}{75}},
  \bibinfo{pages}{1347} (\bibinfo{year}{1981}).

\bibitem[{\citenamefont{Trotter}(1959)}]{trotter__1959}
\bibinfo{author}{\bibfnamefont{H.~F.} \bibnamefont{Trotter}},
  \bibinfo{journal}{Proc. Am. Math Soc.} \textbf{\bibinfo{volume}{10}},
  \bibinfo{pages}{545} (\bibinfo{year}{1959}).

\bibitem[{\citenamefont{Freeman and Doll}(1984)}]{freeman_mboxmonte_1984}
\bibinfo{author}{\bibfnamefont{D.~L.} \bibnamefont{Freeman}} \bibnamefont{and}
  \bibinfo{author}{\bibfnamefont{J.~D.} \bibnamefont{Doll}},
  \bibinfo{journal}{J. Chem. Phys.} \textbf{\bibinfo{volume}{80}},
  \bibinfo{pages}{5709} (\bibinfo{year}{1984}).

\bibitem[{\citenamefont{Freeman and Doll}(1985)}]{freeman_quantum_1985}
\bibinfo{author}{\bibfnamefont{D.~L.} \bibnamefont{Freeman}} \bibnamefont{and}
  \bibinfo{author}{\bibfnamefont{J.~D.} \bibnamefont{Doll}},
  \bibinfo{journal}{J. Chem. Phys.} \textbf{\bibinfo{volume}{82}},
  \bibinfo{pages}{462} (\bibinfo{year}{1985}).

\bibitem[{\citenamefont{Eleftheriou et~al.}(1999)\citenamefont{Eleftheriou,
  Doll, Curotto, and Freeman}}]{eleftheriou_asymptotic_1999-1}
\bibinfo{author}{\bibfnamefont{M.}~\bibnamefont{Eleftheriou}},
  \bibinfo{author}{\bibfnamefont{J.}~\bibnamefont{Doll}},
  \bibinfo{author}{\bibfnamefont{E.}~\bibnamefont{Curotto}}, \bibnamefont{and}
  \bibinfo{author}{\bibfnamefont{D.~L.} \bibnamefont{Freeman}},
  \bibinfo{journal}{J. Chem. Phys.} \textbf{\bibinfo{volume}{110}},
  \bibinfo{pages}{6657} (\bibinfo{year}{1999}).

\bibitem[{\citenamefont{Doll et~al.}(1985)\citenamefont{Doll, Coalson, and
  Freeman}}]{doll_mboxfourier_1985}
\bibinfo{author}{\bibfnamefont{J.}~\bibnamefont{Doll}},
  \bibinfo{author}{\bibfnamefont{R.~D.} \bibnamefont{Coalson}},
  \bibnamefont{and} \bibinfo{author}{\bibfnamefont{D.~L.}
  \bibnamefont{Freeman}}, \bibinfo{journal}{Phys. Rev. Lett.}
  \textbf{\bibinfo{volume}{55}}, \bibinfo{pages}{1} (\bibinfo{year}{1985}).

\bibitem[{\citenamefont{Coalson et~al.}(1986)\citenamefont{Coalson, Freeman,
  and Doll}}]{coalson_partial_1986}
\bibinfo{author}{\bibfnamefont{R.~D.} \bibnamefont{Coalson}},
  \bibinfo{author}{\bibfnamefont{D.~L.} \bibnamefont{Freeman}},
  \bibnamefont{and} \bibinfo{author}{\bibfnamefont{J.~D.} \bibnamefont{Doll}},
  \bibinfo{journal}{J. Chem. Phys.} \textbf{\bibinfo{volume}{85}},
  \bibinfo{pages}{4567} (\bibinfo{year}{1986}).

\bibitem[{\citenamefont{Predescu and Doll}(2002)}]{predescu_optimal_2002}
\bibinfo{author}{\bibfnamefont{C.}~\bibnamefont{Predescu}} \bibnamefont{and}
  \bibinfo{author}{\bibfnamefont{J.~D.} \bibnamefont{Doll}},
  \bibinfo{journal}{J. Chem. Phys.} \textbf{\bibinfo{volume}{117}},
  \bibinfo{pages}{7448} (\bibinfo{year}{2002}).

\bibitem[{\citenamefont{Predescu
  et~al.}(2003{\natexlab{a}})\citenamefont{Predescu, Sabo, and
  Doll}}]{predescu_numerical_2003-1}
\bibinfo{author}{\bibfnamefont{C.}~\bibnamefont{Predescu}},
  \bibinfo{author}{\bibfnamefont{D.}~\bibnamefont{Sabo}}, \bibnamefont{and}
  \bibinfo{author}{\bibfnamefont{J.~D.} \bibnamefont{Doll}},
  \bibinfo{journal}{J. Chem. Phys.} \textbf{\bibinfo{volume}{119}},
  \bibinfo{pages}{4641} (\bibinfo{year}{2003}{\natexlab{a}}).

\bibitem[{\citenamefont{Predescu and Doll}(2003)}]{predescu_random_2003}
\bibinfo{author}{\bibfnamefont{C.}~\bibnamefont{Predescu}} \bibnamefont{and}
  \bibinfo{author}{\bibfnamefont{J.~D.} \bibnamefont{Doll}},
  \bibinfo{journal}{Phys. Rev. E.} \textbf{\bibinfo{volume}{67}},
  \bibinfo{pages}{026124} (\bibinfo{year}{2003}).

\bibitem[{\citenamefont{Predescu
  et~al.}(2003{\natexlab{b}})\citenamefont{Predescu, Sabo, Doll, and
  Freeman}}]{predescu_energy_2003}
\bibinfo{author}{\bibfnamefont{C.}~\bibnamefont{Predescu}},
  \bibinfo{author}{\bibfnamefont{D.}~\bibnamefont{Sabo}},
  \bibinfo{author}{\bibfnamefont{J.~D.} \bibnamefont{Doll}}, \bibnamefont{and}
  \bibinfo{author}{\bibfnamefont{D.~L.} \bibnamefont{Freeman}},
  \bibinfo{journal}{J. Chem. Phys.} \textbf{\bibinfo{volume}{119}},
  \bibinfo{pages}{10475} (\bibinfo{year}{2003}{\natexlab{b}}).

\bibitem[{\citenamefont{Predescu
  et~al.}(2003{\natexlab{c}})\citenamefont{Predescu, Sabo, Doll, and
  Freeman}}]{predescu_heat_2003-1}
\bibinfo{author}{\bibfnamefont{C.}~\bibnamefont{Predescu}},
  \bibinfo{author}{\bibfnamefont{D.}~\bibnamefont{Sabo}},
  \bibinfo{author}{\bibfnamefont{J.~D.} \bibnamefont{Doll}}, \bibnamefont{and}
  \bibinfo{author}{\bibfnamefont{D.~L.} \bibnamefont{Freeman}},
  \bibinfo{journal}{J. Chem. Phys.} \textbf{\bibinfo{volume}{119}},
  \bibinfo{pages}{12119} (\bibinfo{year}{2003}{\natexlab{c}}).

\bibitem[{\citenamefont{Predescu
  et~al.}(2003{\natexlab{d}})\citenamefont{Predescu, Doll, and
  Freeman}}]{predescu_asymptotic_2003}
\bibinfo{author}{\bibfnamefont{C.}~\bibnamefont{Predescu}},
  \bibinfo{author}{\bibfnamefont{J.}~\bibnamefont{Doll}}, \bibnamefont{and}
  \bibinfo{author}{\bibfnamefont{D.~L.} \bibnamefont{Freeman}},
  \bibinfo{journal}{{arXiv:cond-mat/0301525}}
  (\bibinfo{year}{2003}{\natexlab{d}}).

\bibitem[{\citenamefont{Kunikeev et~al.}(2009)\citenamefont{Kunikeev, Freeman,
  and Doll}}]{kunikeev_numerical_2009}
\bibinfo{author}{\bibfnamefont{S.}~\bibnamefont{Kunikeev}},
  \bibinfo{author}{\bibfnamefont{D.~L.} \bibnamefont{Freeman}},
  \bibnamefont{and} \bibinfo{author}{\bibfnamefont{J.}~\bibnamefont{Doll}},
  \bibinfo{journal}{Int. J. Quant. Chem.} \textbf{\bibinfo{volume}{109}},
  \bibinfo{pages}{2916} (\bibinfo{year}{2009}).

\bibitem[{\citenamefont{Singer}(1958)}]{singer_use_1958}
\bibinfo{author}{\bibfnamefont{K.}~\bibnamefont{Singer}},
  \bibinfo{journal}{Nature} \textbf{\bibinfo{volume}{181}},
  \bibinfo{pages}{262} (\bibinfo{year}{1958}).

\bibitem[{\citenamefont{Kubo}(1962)}]{kubo_generalized_1962}
\bibinfo{author}{\bibfnamefont{R.}~\bibnamefont{Kubo}}, \bibinfo{journal}{J.
  Phys. Soc. Japan} \textbf{\bibinfo{volume}{17}}, \bibinfo{pages}{1100–1120}
  (\bibinfo{year}{1962}).

\bibitem[{\citenamefont{Coalson et~al.}(1989)\citenamefont{Coalson, Freeman,
  and Doll}}]{coalson_cumulant_1989}
\bibinfo{author}{\bibfnamefont{R.~D.} \bibnamefont{Coalson}},
  \bibinfo{author}{\bibfnamefont{D.~L.} \bibnamefont{Freeman}},
  \bibnamefont{and} \bibinfo{author}{\bibfnamefont{J.~D.} \bibnamefont{Doll}},
  \bibinfo{journal}{J. Chem. Phys.} \textbf{\bibinfo{volume}{91}},
  \bibinfo{pages}{4242} (\bibinfo{year}{1989}).

\bibitem[{\citenamefont{Press et~al.}(1992)\citenamefont{Press, Teukolsky,
  Vetterling, and Flannery}}]{press_numerical_1992}
\bibinfo{author}{\bibfnamefont{W.~H.} \bibnamefont{Press}},
  \bibinfo{author}{\bibfnamefont{S.~A.} \bibnamefont{Teukolsky}},
  \bibinfo{author}{\bibfnamefont{W.~T.} \bibnamefont{Vetterling}},
  \bibnamefont{and} \bibinfo{author}{\bibfnamefont{B.~P.}
  \bibnamefont{Flannery}}, \emph{\bibinfo{title}{Numerical Recipes in Fortran
  77: The Art of Scientific Computing, Second Edition}}
  (\bibinfo{publisher}{Cambridge University Press}, \bibinfo{address}{New
  York}, \bibinfo{year}{1992}).

\bibitem[{\citenamefont{Golub and Loan}(1996)}]{golub_matrix_1996}
\bibinfo{author}{\bibfnamefont{G.}~\bibnamefont{Golub}} \bibnamefont{and}
  \bibinfo{author}{\bibfnamefont{C.~V.} \bibnamefont{Loan}},
  \emph{\bibinfo{title}{Matrix Computations}} (\bibinfo{publisher}{The John
  Hopkins University Press}, \bibinfo{address}{Baltimore and London},
  \bibinfo{year}{1996}).

\bibitem[{\citenamefont{Englert}(1963)}]{englert_linked_1963}
\bibinfo{author}{\bibfnamefont{F.}~\bibnamefont{Englert}},
  \bibinfo{journal}{Phys. Rev.} \textbf{\bibinfo{volume}{129}},
  \bibinfo{pages}{567} (\bibinfo{year}{1963}).

\bibitem[{\citenamefont{Arai and Goodman}(1967)}]{arai_cumulant_1967}
\bibinfo{author}{\bibfnamefont{T.}~\bibnamefont{Arai}} \bibnamefont{and}
  \bibinfo{author}{\bibfnamefont{B.}~\bibnamefont{Goodman}},
  \bibinfo{journal}{Phys. Rev.} \textbf{\bibinfo{volume}{155}},
  \bibinfo{pages}{514} (\bibinfo{year}{1967}).

\bibitem[{\citenamefont{Fox}(1976)}]{fox_critique_1976}
\bibinfo{author}{\bibfnamefont{R.~F.} \bibnamefont{Fox}}, \bibinfo{journal}{J.
  Math. Phys.} \textbf{\bibinfo{volume}{17}}, \bibinfo{pages}{1148}
  (\bibinfo{year}{1976}).

\bibitem[{\citenamefont{Becker and Brening}(1990)}]{becker_unknown_1990}
\bibinfo{author}{\bibfnamefont{K.}~\bibnamefont{Becker}} \bibnamefont{and}
  \bibinfo{author}{\bibfnamefont{W.}~\bibnamefont{Brening}},
  \bibinfo{journal}{Z. Physik B} \textbf{\bibinfo{volume}{79}},
  \bibinfo{pages}{195} (\bibinfo{year}{1990}).

\bibitem[{\citenamefont{Sanyal et~al.}(1993)\citenamefont{Sanyal, Mandal, Guha,
  and Mukherjee}}]{sanyal_systematic_1993}
\bibinfo{author}{\bibfnamefont{G.}~\bibnamefont{Sanyal}},
  \bibinfo{author}{\bibfnamefont{S.}~\bibnamefont{Mandal}},
  \bibinfo{author}{\bibfnamefont{S.}~\bibnamefont{Guha}}, \bibnamefont{and}
  \bibinfo{author}{\bibfnamefont{D.}~\bibnamefont{Mukherjee}},
  \bibinfo{journal}{Phys. Rev. E.} \textbf{\bibinfo{volume}{48}},
  \bibinfo{pages}{3373–3389} (\bibinfo{year}{1993}).

\bibitem[{\citenamefont{Fauser and Wolter}(1996)}]{fauser_cumulants_1996}
\bibinfo{author}{\bibfnamefont{R.}~\bibnamefont{Fauser}} \bibnamefont{and}
  \bibinfo{author}{\bibfnamefont{H.~H.} \bibnamefont{Wolter}},
  \bibinfo{journal}{Nuclear Physics A} \textbf{\bibinfo{volume}{600}},
  \bibinfo{pages}{491–508} (\bibinfo{year}{1996}).

\bibitem[{\citenamefont{Mandal et~al.}(2002)\citenamefont{Mandal, Ghosh,
  Sanyal, and Mukherjee}}]{mandal_non-perturbative_2002}
\bibinfo{author}{\bibfnamefont{S.~H.} \bibnamefont{Mandal}},
  \bibinfo{author}{\bibfnamefont{R.}~\bibnamefont{Ghosh}},
  \bibinfo{author}{\bibfnamefont{G.}~\bibnamefont{Sanyal}}, \bibnamefont{and}
  \bibinfo{author}{\bibfnamefont{D.}~\bibnamefont{Mukherjee}},
  \bibinfo{journal}{Chem. Phys. Lett.} \textbf{\bibinfo{volume}{352}},
  \bibinfo{pages}{63–69} (\bibinfo{year}{2002}).

\bibitem[{\citenamefont{Negele and Orland}(1987)}]{negele_quantum_1987}
\bibinfo{author}{\bibfnamefont{J.}~\bibnamefont{Negele}} \bibnamefont{and}
  \bibinfo{author}{\bibfnamefont{H.}~\bibnamefont{Orland}},
  \emph{\bibinfo{title}{Quantum Many-particle Systems}}
  (\bibinfo{publisher}{{Addison-Wesley}}, \bibinfo{address}{Redwood City},
  \bibinfo{year}{1987}).

\bibitem[{\citenamefont{Goldenfeld}(1982)}]{goldenfeld_lecturesphase_1982}
\bibinfo{author}{\bibfnamefont{N.}~\bibnamefont{Goldenfeld}},
  \emph{\bibinfo{title}{Lectures on phase transitions and the renormalization
  group}} (\bibinfo{publisher}{{Addison-Wesley}}, \bibinfo{address}{Reading,
  {MA}}, \bibinfo{year}{1982}).

\bibitem[{\citenamefont{Abramowitz and
  Stegun}(1965)}]{abramowitz_handbook_1965}
\bibinfo{author}{\bibfnamefont{M.}~\bibnamefont{Abramowitz}} \bibnamefont{and}
  \bibinfo{author}{\bibfnamefont{I.}~\bibnamefont{Stegun}},
  \emph{\bibinfo{title}{Handbook of Mathematical Functions}}
  (\bibinfo{publisher}{Dover}, \bibinfo{address}{New York},
  \bibinfo{year}{1965}).

\end{thebibliography}
\newpage
\section*{Figure Captions}
\begin{enumerate}
\item \label{fig:one}
Graphical representation of set partitions that provide the coefficients in the cumulant expansion.  The top diagram represents the improper partition $S=(1,2,3,4)$.  The second line represents all ways of partitioning three elements into one group with the fourth element remaining.  The third line shows all possible pair partitions, the fourth and fifth lines represent partitions containing a single pair, and the final line shows the last possibility consisting of four clusters each with a single vertex.  Each of the lines correspond in order to the 5 terms in Eq.(\ref{1.8.1}) with the corresponding product of cumulants and associated coefficients indicated.
\item \label{fig:two}
Diagrammatic representation of the cumulants.  Each solid circle represents a vertex function $V_{PA}(i)$ and each line represents an interaction $\Gamma_{ij}$ (see text for definitions).  In (a) the contributions to the second-order cumulant is shown, with the first diagram in (a) being unlinked and contributing 0.  Contributions to the third-order cumulant are shown in (b), with the first three diagrams representing terms in the second line of Eq. (\ref{1.33}) and the last diagram being an example of a loop diagram.  In (c) three fourth-order diagrams are shown demonstrating consecutive, centered and loop diagrams respectively.
\end{enumerate}
\newpage
\begin{figure}
\includegraphics[clip=true,width=16cm]{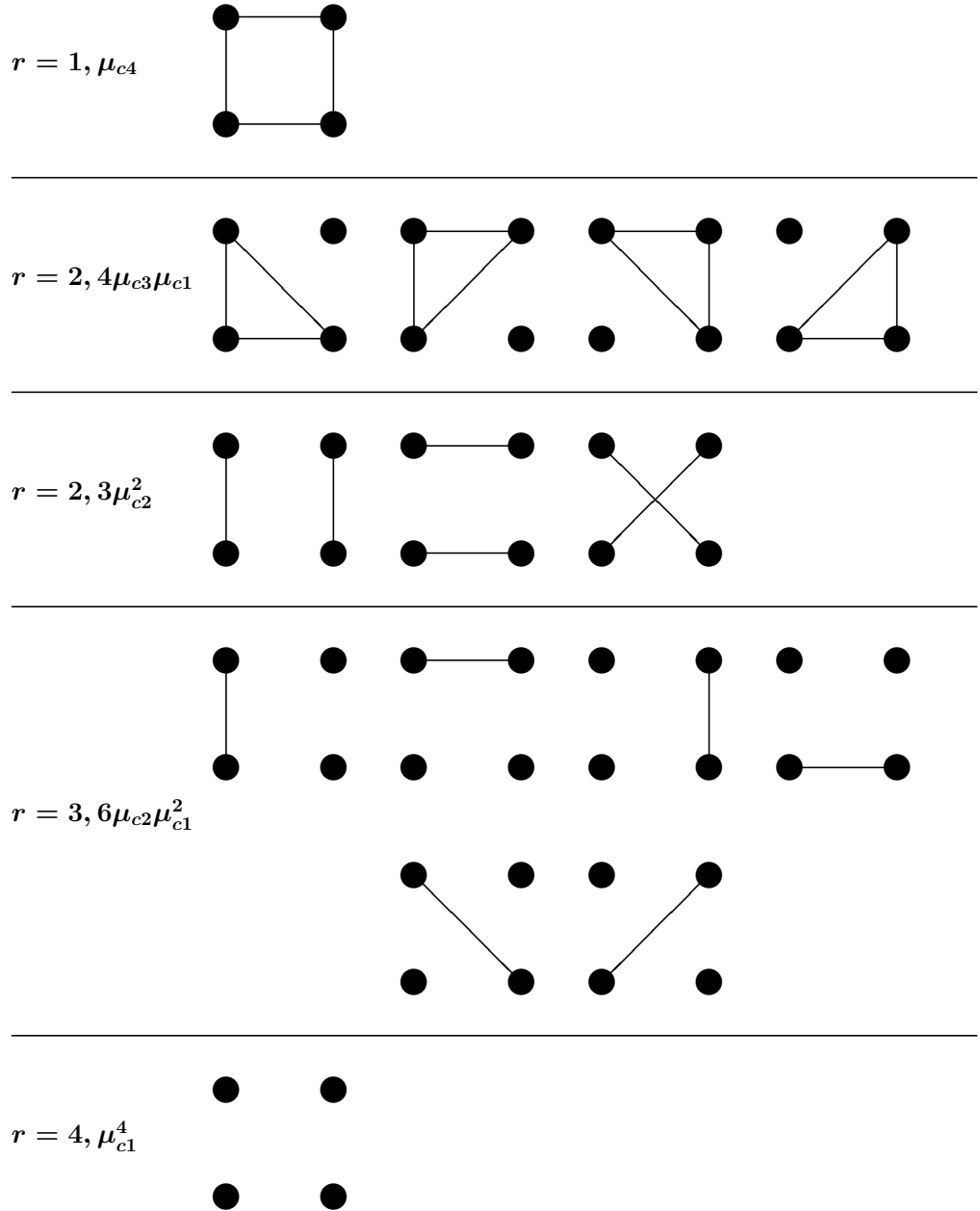}
\caption{Kunikeev {\em et al.}}
\end{figure}
\newpage
\begin{figure}
\includegraphics[clip=true,width=16cm]{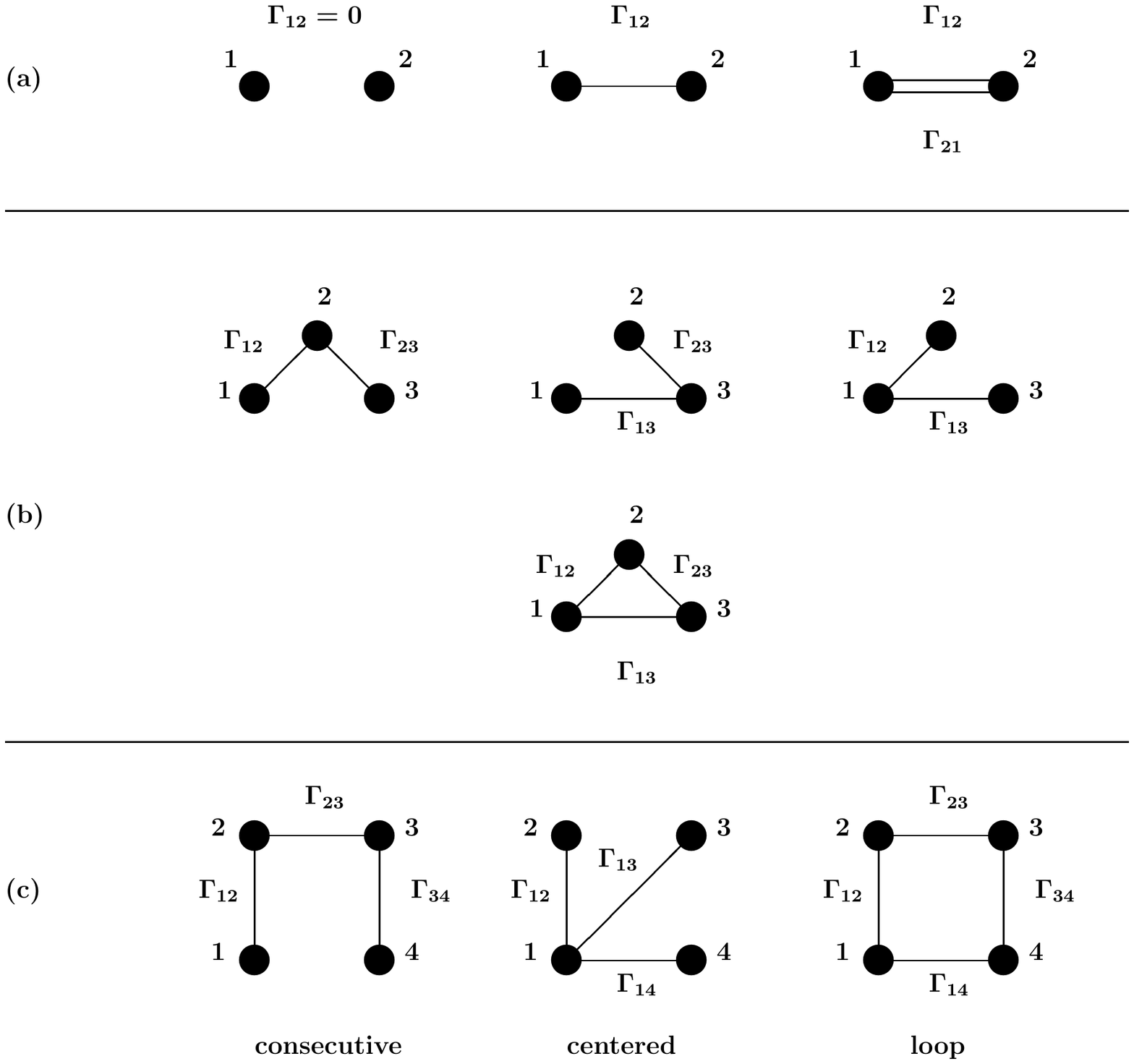}
\caption{Kunikeev {\em et al.}}
\end{figure}
%\newpage
%\begin{figure}
%\includegraphics[clip=true,width=20cm]{Fig_3a.pdf}
%\caption{Kunikeev {\em et al.}}
%\end{figure}

\end{document}